\documentclass[12pt]{article}
\usepackage{amsmath}
\usepackage{amsthm}
\usepackage{color}
\usepackage{amsfonts}
\usepackage{latexsym}
\usepackage{amssymb}
\usepackage{graphicx}
\usepackage{mathrsfs}
\usepackage{bbm}
\usepackage[page]{appendix} 

\newif{\ifcomentarios}
\comentariosfalse

\newtheorem{Thm}{Theorem}[section]
\newtheorem{theorem}[Thm]{Theorem}
\newtheorem{proposition}[Thm]{Proposition}
\newtheorem{corollary}[Thm]{Corollary}
\newtheorem{lemma}[Thm]{Lemma}

\newtheorem{remark}{Remark}[section]

\newtheorem{definition}[Thm]{Definition}

\newcommand{\dsum}{\displaystyle \sum}
\renewcommand{\mathcal}{\mathscr}
\newcommand\mfrac[2]{\genfrac{}{}{0pt}{}{#1}{#2}}

\DeclareMathOperator*{\slim}{s-lim}
\numberwithin{equation}{subsection}
\let\oldsection\section\renewcommand{\section}{  \renewcommand{\theequation}{\thesection.\arabic{equation}}  \renewcommand{\thetheorem}{\thesection.\arabic{Thm}}
  \oldsection}
\let\oldsubsection\subsection\renewcommand{\subsection}{  \renewcommand{\theequation}{\thesubsection.\arabic{equation}}  \renewcommand{\thetheorem}{\thesubsection.\arabic{Thm}} 
  \oldsubsection}

\begin{document}

\title{A mathematical theory of the critical point \\
of ferromagnetic Ising systems}
\author{Domingos H. U. Marchetti\thanks{%
marchett@if.usp.br, Instituto de Fisica, Universidade de S\~ao Paulo (USP),
Brasil}, Manfred
Requardt\thanks{requardt@theorie.physik.uni-goettingen.de, Institut
f\"ur Theoretische Physik, G\"ottingen} \\
and Walter F. Wreszinski\thanks{%
wreszins@gmail.com, Instituto de Fisica, Universidade de S\~ao Paulo (USP),
Brasil}}
\date{9-4-2024}
\maketitle

\tableofcontents

\begin{abstract}
We develop a theory of the critical point of the ferromagnetic Ising model,
whose basic objects are the ergodic (pure) states of the infinite system. It
proves the existence of anomalous critical fluctuations, for dimension $
\nu=2 $ and, under a standard assumption, for $\nu=3$, for the model with
nearest-neighbor interaction, in a way which is consistent with the
probabilistic approach of Cassandro, Jona-Lasinio, and several others,
reviewed in Jona-Lasinio's article in Phys. Rep. 352,439 (2001). We propose
to single out the state at the critical temperature $T_{c}$, among the
ergodic thermal states associated to temperatures $0\le T\le T_{c}$, by a
condition of non-summable clustering of the connected two-point function.
The analogous condition on the connected (2r)- point functions, for $r \ge 2$
, together with a scaling hypothesis, natural within our framework, proves that 
the (macroscopic) fluctuation state is quasi-free, after a proper rescaling, 
also at the critical temperature, in agreement with a theorem by Cassandro
and Jona-Lasinio, whose proof is, however, shown to be incomplete.
Other subjects treated include topics relating to universality, including
spontaneous breaking of continuous symmetries and violations of universality
in problems of energetic and dynamic stability.
\end{abstract}

\medskip


\section{General overview on phase transitions, of direct relevance to the
present paper}

\setcounter{equation}{0} \setcounter{Thm}{0}

\subsection{Introduction and summary}

\label{1.1}

As remarked by Glimm in his review of certain books and articles on
statistical mechanics \cite{Gli}, \textquotedblleft statistical mechanics
concerns infinite systems, and the relevant analysis is over
infinite-dimensional spaces\textquotedblright . We shall use his
illuminating introdution as a useful guide. Let $\Lambda $ a finite subset of 
$\mathbb{Z}^{\nu }$, with volume (identified with the number of sites) $
|\Lambda |$, with $\nu$ the dimension and $\eta (\Lambda )$ denote
the set of nearest-neighbor pairs $x,y\in \Lambda $ with $\left\vert
x-y\right\vert =1$. Consider the measures on $\mathbb{R}^{\left\vert \Lambda
\right\vert }$ 
\begin{equation}
d\mu _{T,h}^{\Lambda }\equiv \prod_{x\in \Lambda }\exp (h\sigma
_{x}/T)\prod_{x,y\in \eta (\Lambda )}\exp (-J\sigma _{x}\sigma _{y}/T)d\mu
_{\Lambda }  \label{(1.1a)}
\end{equation}
where 
\begin{equation}
d\mu _{\Lambda }\equiv \prod_{x\in \Lambda }d\mu (\sigma _{x})\qquad \text{
with}\qquad \text{ }d\mu (\sigma _{x})=\frac{\delta _{1}(\sigma _{x})+\delta
_{-1}(\sigma _{x})}{2}d\sigma _{x}  \label{(1.1b)}
\end{equation}
and $\delta _{a}$ denotes the Dirac delta measure concentrated at $\alpha $
so $\sigma _{x}\in \left\{ {-1,1}\right\} $ are discrete Ising variables. $h$
and $J$ denote the real values of the external field and the couplings, and $
T$ the absolute temperature (the Boltzmann constant $k$ being set equal to
one). $d\mu _{\Lambda }$ defines the model of Ising spins in a finite box $
\Lambda $. The free energy $f_{\Lambda }(T,h)$ is defined by 
\begin{equation}
f_{\Lambda }(T,h)\equiv -T\frac{1}{|\Lambda |}\log \int d\mu _{T,h}^{\Lambda
}  \label{(1.2)}
\end{equation}
The thermodynamic limit $\Lambda \nearrow \mathbb{Z}^{\nu }$ (in the simple
case of $\Lambda $ a cubic box of side $a$, it is the limit $a\rightarrow
\infty $) exists \cite{CJT}: 
\begin{equation}
f(T,h)\equiv \lim_{\Lambda \nearrow \mathbb{Z}^{\nu }}f_{\Lambda }(T,h)
\label{(1.3)}
\end{equation}
While $f_{\Lambda }(T,h)$ is an analytic function of $T$ and $h$ for any $
\Lambda $ with $\left\vert \Lambda \right\vert <\infty $, $f(T,h)$ is only
piecewise analytic, and the boundaries between the analytic pieces of $
f(T,h) $ correspond to \emph{phase transitions}. In the lattice-gas picture
with variables $n_{x}=\left( 1+\sigma _{x}\right) /2$, with $n_{x}=0\text{
or }1$ according to whether the site $x$ is occupied or empty, it describes,
for example, a liquid-vapor transition. In the original variables, there
occurs a \emph{ferromagnetic phase transition}.

As regards the description of real systems, the Ising model in dimension $
\nu=3$ is one of the most successful models of both ferromagnets and fluids,
due to the fact that most magnetic materials are (highly) anisotropic
\cite{PV}, and the lattice-gas analogy is very effective at the
critical point \cite{SS}.

The measure $d\mu _{T,h}^{\Lambda }$ is not completely determined by the
r.h.s. of equation \eqref{(1.1a)}: we need to impose the boundary conditions
(b.c.), where the spins $\sigma _{x}$ at the boundary: 
\begin{equation}
\ \delta \Lambda \equiv \{x:\text{dist}(x,\Lambda )=1\}  \label{(1.4)}
\end{equation}
are specified in \eqref{(1.1a)} with the nearest-neighbors set $\eta
(\Lambda )$ replaced by $\eta \left( \Lambda \cup \delta \Lambda \right) $.
For high temperature $T$ in \eqref{(1.1a)},\eqref{(1.1b)} (\textquotedblleft
weak couplings\textquotedblright\ $J/T<<1$), $h=0$, or any value of $J$ if $h \ne 0$), 
\begin{equation}
d\mu _{T,h}\equiv \lim_{\Lambda \nearrow \mathbb{Z}^{\nu }}d\mu
_{T,h}^{\Lambda }  \label{(1.5)}
\end{equation}
is independent of the b.c., but for low temperature $T$ and $h=0$
(\textquotedblleft strong coupling\textquotedblright ), and $\nu \geq 2$, it
is not: this phenomenon is a second notion of a phase transition. A phase
transition also occurs in the previously mentioned sense (to which we refer
for definiteness as \textquotedblleft first notion\textquotedblright ) of
non-piecewise analyticity. More precisely, $\partial f/\partial h$ has a
jump discontinuity (called spontaneous magnetization $m_{s}$) at $h=0$: 
\begin{equation}
m_{s}\equiv \lim_{h\rightarrow 0+}\frac{\partial f}{\partial h}
\label{(1.6)}
\end{equation}
Note that the second notion of phase transition is different from the first
one, because $f(T,h)$ is independent of the b.c. \cite{CJT}. 

Note also that
in \eqref{(1.6)}, the condition $m_{s} >0$ characterizes a first-order 
transition, which occurs for $0 \le T <T_{c}$, but, at the critical point, 
$m_{s} = m_{s}(T) \to 0$ as $T \to T_{c}^{-}$ (also as $T \to T_{c}^{+}$)
continuously, see the forthcoming Remark \ref{Remark 1.3.2} and the reference
given there. Near $T=T_{c}$, we have the forthcoming (1.1.17), i.e.,
the susceptibility diverges, and therefore the critical point is a second-order
point. This is further confirmed by looking at the manifold of points $(T,h)$, 
the specific heat diverges at $T=T_{c}$ when $h=0$. 

In this paper we restrict ourselves for simplicity to the situation of zero
external field ($h=0$): the only free parameter is, therefore, the
temperature $T$, which is taken to vary in the range $0 \le T \le T_{c}$, in
which phase transitions occur.

As remarked by Glimm, \textquotedblleft Since $d\mu _{T}$ is defined by the
limit $\Lambda \nearrow \mathbb{Z}^{\nu }$, $\delta \Lambda \rightarrow
\infty $, weak coupling leads to (exact) independence of $\delta \Lambda $
and $d\mu _{T}$, i.e., independence of random variables (r.v.) separated by
an infinite distance, while strong coupling allows phase transition and
dependence between r.v. separated by an infinite distance\textquotedblright .

In the last half-century, this idea was considerably developed, perhaps most
deeply, in the realm of the Ising model, through the probabilistic approach
to the theory of the renormalization group (RG), championned by several
people, among them M. Cassandro, G. Jona-Lasinio, G. Gallavotti,
A. Martin-L\"{o}f, P. M. Bleher and Ya. G. Sinai (\cite{MCJL1},
\cite{MCJL2}, \cite{GML}, \cite{BS1}, \cite{BS2}), and reviewed by
Jona-Lasinio in \cite{JL}, to 
which we refer for numerous additional references. A basic result in this
context was Bleher and Sinai's treatment \cite{BS1}, \cite{BS2} of the
critical behavior of Dyson's hierarchical model \cite{Dy}, which was
expected to describe qualitatively well the critical behavior of the Ising
model with both short-range and long-range interactions. These seminal
papers greatly stimulated further work on hierachical models for classical
spin systems, see, in particular, \cite{MCG} and references therein.

The measure $d\mu _{T}^{\Lambda }=d\mu _{T,h=0}^{\Lambda }$ generates a
state $\omega _{T}^{\Lambda }$ over the (abelian) $\mathrm{C}^{\ast
}$-algebra of functions $f_{\Lambda }$ of the variables $\{\sigma
_{x};\ x\in \Lambda \}$ by 
\begin{equation}
\omega _{T}^{\Lambda }(f_{\Lambda })\equiv \frac{\displaystyle\int
f_{\Lambda }(\sigma _{x})d\mu _{T}^{\Lambda }(\sigma _{x})}{\displaystyle 
\int d\mu _{T}^{\Lambda }(\sigma _{x})}  \label{(1.6a)}
\end{equation}
This is a positive, normalized functional of the $f_{\Lambda }$: 
\begin{equation}
\omega _{T}^{\Lambda }(f_{\Lambda })\geq 0\qquad \text{ if } \qquad
f_{\Lambda }\geq 0 \quad \text{ and } \quad \omega _{T}^{\Lambda }(\mathbf{1})=1
\label{(1.7)}
\end{equation}
The limit measure, the limit $d\mu $ as $\Lambda \nearrow \mathbb{Z}^{\nu }$,
$\delta \Lambda \rightarrow \infty $, of the $d\mu _{T}^{\Lambda }$, is
associated to a \emph{state of the infinite system} $\omega _{T}$ by 
\begin{equation}
\omega _{T}(f)=\!``\lim_{\Lambda \nearrow \mathbb{Z}^{\nu }}\!"\omega
_{T}^{\Lambda }(f_{\Lambda })  \label{(1.9)}
\end{equation}
where the quotation marks above mean that $f$ is the limit of
$f_{\Lambda }$ in the sup norm (\cite{RR}, Section 1). We may also
view (this is the natural approach from a physical standpoint) the
Ising model as an anisotropic limit of a quantum model: in this case
(the forthcoming  \eqref{(1.2.1)}, one may speak of the
\textquotedblleft Ising universality class\textquotedblright , for
which phase transitions with spontaneous symmetry breaking (s.s.b.) of
a discrete symmetry, the spin-flip symmetry $ \sigma _{x}\rightarrow
-\sigma _{x},  ~\forall x\in \mathbb{Z}^{\nu }$ (and its generalization in
the quantum case) take place, see also Subection \ref{1.3.1}. The
finite-region algebras $\mathcal{A}_{\Lambda }$ are then the 
spin algebras generated by the Pauli matrices $\{\sigma _{x}^{1,2,3}\}$, and
the \textquotedblleft quasilocal algebra\textquotedblright\ $\mathcal{A}$ is
the inductive limit of the $\mathcal{A}_{\Lambda }$ (i.e., the norm closure
of the union of the $\mathcal{A}_{\Lambda }$, see Subsection \ref{1.2}. We
adopt this view (which will be particularly useful for the dynamics in
Subsection 3.2; restriction to the abelian algebra generated by the $
\{\sigma _{x}^{3}\}$ coincides with the previous description.

The study of statistical mechanics from the point of view of the states of
infinite systems in both classical and quantum systems has a long history, beginning with 
the paper of Haag, Hugenholtz and Winnink \cite{HaHW}, and the papers of Lanford
and Ruelle \cite{LRu} (see also Ruelle's book \cite{Rue}), and going over to the
results of uniqueness of the equilibrium state of the Ising model for $T>T_{c}$ due to 
Dobrushin \cite{Dob} and Lebowitz and Martin-L\"{o}f \cite{LML}. The books by Simon \cite{Sim}
and Israel \cite{Isr} are good sources from the point of view of treating classical
and quantum models in parallel. 

The dependence vs independence between random variables separated by an
infinite distance, which leads to the \textquotedblleft sharp dichotomy
between strong and weak coupling\textquotedblright\ mentioned by Glimm
(\cite{Gli}, p. 675), \textquotedblleft can only exist for infinite
systems: in 
finite systems there is no 'infinitely distant separation' between the
degrees of freedom, i.e., between the coordinate directions in the
finite-dimensional space of variables defining the problem\textquotedblright
. A natural way to define this dependence precisely is by the rate of
clustering of the (necessarily!) states of the \emph{infinite} system $%
\omega _{T}$. If the latter are assumed to be translation-invariant, i.e.,
invariant under the translation $\tau _{x}$ (defined by $\tau _{x}(\sigma
_{y}^{3})=\sigma _{x+y}^{3}$), we shall see in Subsection \ref{1.2} that
these are \textquotedblleft automorphisms of $\mathcal{A}$), this clustering
may be expressed by the property (the capital letter \textquotedblleft
C\textquotedblright\ below refers to \textquotedblleft
connected\textquotedblright\ rather than \textquotedblleft
critical\textquotedblright ): $\forall A,B\in \mathcal{A}$ we have 
\begin{equation}
W_{\omega _{T},C}^{\mathrm{e}}(A,B;x)\equiv \omega _{T}^{\mathrm{e}}(A\tau
_{x}(B))-\omega ^{\mathrm{e}}(A)\omega ^{\mathrm{e}}(B)\longrightarrow 0 
\text{ \qquad as\qquad\ }\left\vert x\right\vert \longrightarrow \infty
\label{(1.10)}
\end{equation}
The capital letter $W$ above refers to \textquotedblleft
Wightman\textquotedblright\ in order to emphasize the close connection with
quantum field theory, see Subsection \ref{1.4}. Other terminologies in common use
are ``cumulants'', ``truncated correlatin functions'', and ``Ursell functions''.
Equation \eqref{(1.10)} is a
general property of a special class of states, namely those ergodic states
$\omega ^{\mathrm{e}}$, which are also in the subset of the set of extremal states
which are invariant under a group (see, e.g. \cite{Rue}, Definition 6.3.1). 
In our case, the group is that of lattice translations on $\mathbf{Z}^{\nu}$,
and the extremal states are identified with the pure
phases \cite{Wight}: they cannot be written as a convex combination of
two other states $\omega _{1}\neq \omega ^{\mathrm{e}}$ and $\omega
_{2}\neq \omega ^{\mathrm{e}}$, i.e., the relation $\omega
^{\mathrm{e}}=a\omega _{1}+(1-a)\omega _{2}
\text{ with }0<a<1$ does \emph{not} hold unless $\omega _{1}=\omega ^{
\mathrm{e}}$ and $\omega _{2}=\omega ^{\mathrm{e}}$. We now restrict our
discussion to the phase-transition region of the Ising model, the extremal
ergodic states being $\omega _{1}^{\mathrm{e}}=\omega _{T}^{+}$ and $\omega
_{2}^{\mathrm{e}}=\omega _{T}^{-}$; these states may be obtained as limits
(as previously described) of the states $\omega _{T}^{\Lambda }$, as $
\Lambda \nearrow \mathbb{Z}^{\nu }$, with $+$ or $-$ b.c., respectively, and
are not invariant under the spin-flip symmetry, i.e., as stated before,
there is s.s.b.. If $0\leq T<T_{c}$, the closed segment 
\begin{equation}
\omega _{T}=a\omega _{T}^{+}+(1-a)\omega _{T}^{-}\qquad \text{ with }\qquad
\ 0\leq a\leq 1  \label{(1.11)}
\end{equation}
describes, geometrically, the two-phase region, with boundary points the
extremal ergodic states (pure phases) - in the lattice-gas language, the
liquid phase and the vapour phase. The critical state $\omega _{T_{c}}$ is
spin-flip invariant and unique, hence also ergodic - it may be obtained as
the limit of the states $\omega _{T_{c}}^{\Lambda }$, with free or periodic
b.c. on $\Lambda $.

\begin{remark}
\label{Remark 1.1}

It should be emphasized that the concepts of extremal and ergodic states
may be distinct, as, for instance, in the Ising antiferromagnet, for which
there are two invariance groups, corresponding to the translations within
each sublattice. In this paper we shall, however, restrict ourselves to the 
Ising universality class, for which the two concepts coincide, and use 
the words ``extremal states'' and ``ergodic states'' interchangeably.

\end{remark}

We now come back to the precise definition of the dependence of the r.v.
mentioned by Glimm. The probabilistic approach to the RG leads to conjecture
that: a.) if $0<T<T_{c}$, the rate of clustering of the connected (Wightman)
two-point functions for the ergodic states $\omega _{T,C}^{\pm }$ given by 
\eqref{(1.10)} is such that 
\begin{equation}
W_{\omega _{T}^{\pm },C}(A,B;x)\in l^{1}(\mathbb{Z}^{\nu })\qquad \text{ if }
\qquad 0\leq T<T_{c}  \label{(1.12)}
\end{equation}
and b.) if $T=T_{c}$, i.e., at the critical point, the clustering is not
summable: 
\begin{equation}
W_{\omega _{T_{c}},C}(A,B;x)\notin l^{1}(\mathbb{Z}^{\nu })  \label{(1.13)}
\end{equation}
Thus, the critical point is singled out among the ergodic states by the
condition of non-summable clustering \eqref{(1.13)}. In the one-dimensional
$1/r^{2}$ model Imbrie and Newman (\cite{IN}, see also \cite{DHUM}) have
proved, however, that there is an intermediate regime $T_{0} < T <T_{c}$ with
positive magnetization but nonsummable connected two-point functions.

\begin{remark}
\label{Remark 1.2}

Although not for Ising models, it is possible that there is a first-order transition,
as happens for n.n. high-q Potts models. At the ``critical'' temperature separating the low-T
and high-T regime, there coexist q extremal (ergodic) ordered states with one ergodic (extremal)
disordered state, all $q+1$ states with exponentially decaying correlations. We thank the referee
for this remark.

\end{remark}

We define the (finite region) susceptibility 
\begin{equation}
\chi _{\Lambda }(T)=\frac{\partial ^{2}f_{\Lambda }}{\partial h^{2}}(T,0)
\label{(1.14)}
\end{equation}
By \eqref{(1.2)}, this is a normalized fluctuation of the r.v. $\sigma
_{x}^{3}$ , i.e., 
\begin{equation}
\chi _{\Lambda }(T)=\omega _{T}^{\Lambda }\left( \left( \left.
\sum\nolimits_{x\in \Lambda }\left( \sigma _{x}^{3}-\omega _{T}^{\Lambda
}(\sigma _{x}^{3})\right) \right/ |\Lambda |^{1/2}\right) ^{2}\right)
\label{(1.15)}
\end{equation}
Near the critical point we have \cite{CJT} 
\begin{equation}
\chi (T)\equiv \lim_{\Lambda \nearrow \mathbb{Z}^{\nu }}\chi _{\Lambda
}(T)\approx (T_{c}-T)^{-\gamma }\qquad \text{ as}\qquad T\nearrow T_{c}
\label{(1.16)}
\end{equation}
Equation \eqref{(1.16)} defines the \emph{critical exponent} $\gamma $; at
the same time, the spontaneous magnetization \eqref{(1.6)} has the behavior 
\begin{equation}
m_{s}\approx (T_{c}-T)^{\beta }\qquad \text{ as}\qquad T\nearrow T_{c}
\label{(1.17)}
\end{equation}
Equation \eqref{(1.17)} defines the critical exponent $\beta $. Equations 
\eqref{(1.16)} and \eqref{(1.17)} lead us to define the normalized
fluctuation operator associated to the r.v. (operator) $\sigma ^{3}$ by the
equation 
\begin{equation}
\tilde{F}_{\Lambda ,T}(\sigma ^{3})\equiv \frac{1}{(\tilde{V}_{T}^{\Lambda
})^{1/2}}\sum_{x\in \Lambda }(\sigma _{x}^{3}-\omega _{\Lambda }^{T}(\sigma
_{x}^{3}))  \label{(1.18)}
\end{equation}
where 
\begin{equation}
\tilde{V}_{T}^{\Lambda }\equiv \omega _{T}^{\Lambda }\left( \left(
\sum\nolimits_{x\in \Lambda }\left( \sigma _{x}^{3}-\omega _{T}^{\Lambda
}(\sigma _{x}^{3})\right) \right) ^{2}\right)  \label{(1.19)}
\end{equation}
is the variance of the \textquotedblleft block-variable\textquotedblright\ $
\sum_{x\in \Lambda }\sigma _{x}^{3}$ in the state $\omega _{\Lambda }^{T}$.
For $0<T<T_{c}$, this is consistent with \eqref{(1.12)} and the
normalization by $\sqrt{\left\vert \Lambda \right\vert })$ in \eqref{(1.15)},
but, for $T=T_{c}$, one expects \emph{\ anomalous fluctuations}, defined by: 
\begin{equation}
V_{\Lambda }^{T}=O(|\Lambda |^{\alpha })\text{ \qquad with\qquad\ }\alpha > 
\frac{1}{2}\text{ \quad as \quad\ }\left\vert \Lambda \right\vert
\longrightarrow \infty  \label{(1.20)}
\end{equation}
These anomalous fluctuations are associated to strong density fluctuations,
whereby light is strongly scattered at the critical point: this is the
beautiful phenomenon of \emph{critical opalescence}, of which one of the
most detailed theoretical discussions is Einstein's early paper \cite{Ein}.
The Ising model, in its lattice-gas version for liquids, is the foremost
model in this connection (\cite{PV}, p. 552), providing ample empirical
support to \eqref{(1.20)} \cite{Fis}.

A different definition of normalized fluctuation operator is

\begin{equation}
F_{\Lambda ,T}(\sigma ^{3})\equiv \frac{1}{\left( V_{\Lambda }^{T}\right)
^{1/2}}\sum_{x\in \Lambda }(\sigma _{x}^{3}-\omega _{T}(\sigma _{0}^{3}))
\label{(1.21)}
\end{equation}
where 
\begin{equation}
V_{\Lambda }^{T}\equiv \omega _{T}\left( \left( \sum\nolimits_{x\in \Lambda
}\left( \sigma _{x}^{3}-\omega _{T}(\sigma _{0}^{3})\right) \right)
^{2}\right)  \label{(1.22)}
\end{equation}

Associated to $F_{\Lambda ,T}(\sigma ^{3})$ and $\tilde{F}_{\Lambda
,T}(\sigma ^{3})$ we have two types of \emph{characteristic functions} $\Phi
_{\Lambda ,T},\tilde{\Phi}_{\Lambda ,T}$, defined by (for $\omega _{T}$ an
ergodic state): 
\begin{equation}
\Phi _{\Lambda ,T}(t)\equiv \omega _{T}(\exp (itF_{\Lambda ,T}))\ ,\qquad
\forall t\in \mathbb{R}  \label{(1.23)}
\end{equation}
and 
\begin{equation}
\tilde{\Phi}_{\Lambda ,T}(t)\equiv \omega _{T}^{\Lambda }((\exp (it\tilde{F}
_{\Lambda ,T}))\ ,\qquad \forall t\in \mathbb{R}  \label{(1.24)}
\end{equation}
The limits (for $\omega _{T}$ an ergodic state) 
\begin{equation}
\Phi _{T}(t)\equiv \lim_{\Lambda \nearrow \mathbb{Z}^{\nu }}\Phi _{\Lambda
,T}(t)\ ,\qquad \forall t\in \mathbb{R}  \label{(1.25)}
\end{equation}
and 
\begin{equation}
\tilde{\Phi}_{T}(t)\equiv \lim_{\Lambda \nearrow \mathbb{Z}^{\nu }}\tilde{
\Phi}_{\Lambda ,T}(t)\ ,\qquad \forall t\in \mathbb{R}  \label{(1.26)}
\end{equation}
(provided they exist) need not coincide, see Subsection \ref{1.3.3}.
Moreover, for instance, the limit \eqref{(1.25)} need not exist in general:
for instance, taking random b.c. on $\Lambda $ on the r.h.s. of \eqref{(1.25)},
for $T<T_{c}$, one may have several distinct
subsequential limits along sequences of finite regions $\Lambda
_{n},n=1.2,...$ of the r.h.s. of \eqref{(1.25)}, see \cite{vES}.

We now finish the summary of contents of our paper.

Since we shall base most of our discussion on the states of infinite
systems, which are the generalizations of \eqref{(1.9)} to the whole
quasi-local algebra $\mathcal{A}$, we make an interlude, in Subsection
\ref{1.2}, in order to establish some of the fundamental notions. 

Subsection \ref{1.3} is concerned with phase transitions and spontaneous
symmetry breaking (s.s.b.) in ferromagnetic spin systems. It is divided into
four subsections.

Subsection \ref{1.3.1} deals with generalities. Subsection \ref{1.3.2} is
concerned with the statement and comments on a criterion of criticality
(Criterion A) based on \eqref{(1.13)}. Subsection \ref{1.3.3} defines SLRO,
LLRO and discusses their connection with Criterion A. Subsection \ref{1.3.4}
introduces the connected (or truncated) $r$- point Wightman functions (also
called Ursell functions) and some of the associated inequalities which are
used in the text. Subsection \ref{1.3.5} defines the concepts of interacting
and quasi-free theories.

It is only in Subsection \ref{1.4} that the main subject of this paper, that
of the relation between macroscopic phenomena and the laws of microphysics,
is introduced. It defines the \emph{fluctuation state} $\omega _{T}^{F}$,
the \textquotedblleft state induced by the fluctuations\textquotedblright\
from an equilibrium state at temperature $T$, together with our main related
assumptions, and its presumed connection with renormalization-group (RG)
transformations. The possible extension of $\omega _{T}^{F}$ to the general
\textquotedblleft quantum Ising systems\textquotedblright , which constitute
the \textquotedblleft Ising universality class\textquotedblright\ associated
to the Ising model, defined in Subsection \ref{1.2}, and the associated
\textquotedblleft Lie algebra of fluctuations\textquotedblright\ (\cite{Ver},
\cite{VZ}, \cite{HL}, \cite{Sew1}) is also briefly discussed there.

In Subsection \ref{1.5} we study the ferromagnetic Ising model in finite
volume. It is shown, as direct corollaries of the work of De Coninck and
Gottal \cite{dCG} that the characteristic function $\Phi _{\Lambda ,T}(t)$
defines an interacting system (equivalently: is Non-Gaussian) for all
temperatures $0\leq T<\infty $. One further result in this framework,
Proposition~\ref{prop:1.5.2}, is used in the proof of Theorem \ref{th:2.3.1}.

Section \ref{2} is devoted to the connected Wightman functions ($r\geq 2$),
both for $0<T<T_{c}$ and for $T=T_{c}$. Subsection \ref{2.1} describes the
framework which was introduced by one of us (M.R.) in \cite{MR} and reviews
the case $0<T<T_{c}$, both for $r=2$ and $r>2$. In Subsection \ref{2.2} we
restrict ourselves to $r=2$ and $T=T_{c}$, establishing the existence of
anomalous fluctuations for $\nu =2$ and for $\nu =3$, under the standard
assumption \eqref{(2.2.22)}. 

In Subsection \ref{2.3} we approach the
question of the connected $r$-point function for $r\geq 4$ in the critical
case. Under an assumption (Scaling Assumption C), which is the natural
analogue of \eqref{(2.2.12b)} for $r\geq 4$, we prove in Theorem
\ref{th:2.3.1} that the fluctuation state $\omega _{F}^{T}$ is
quasi-free (corresponding to a Gaussian characteristic function).
This conjecture seems to be widespread among several mathematical physicists
(private communications by S. Molchanov and G. Slade to one of us (D.H.U.M.)),
but the sole theorem we were able to found which claims to prove it was
Theorem 6.2 in the paper by Cassandro and Jona-Lasinio \cite{MCJL2}. Their
proof is, however, incomplete (see also ~\ref{Remark 1.4.1} and ~\ref{Remark 1.5.1}
,in particular concerning the possible connection with one of the peculiarities
pointed out by Griffiths and Pearce, and the important work of Wehrl).

In Section \ref{3} we study several aspects of universality and its
violation. Using the Dyson model as a
prototype the known universality of critical exponents such as $\beta $ and $
\gamma $ for long-range interactions, is briefly reviewed in
Subsection 3.1. In Subsection 3.2 we argue that
certain properties of 
energetic and dynamical stability, which distinguish long-range interactions
from short-range ones, are among those which violate universality. In order
to prove this conjecture in an explicit model (the Dyson model), we show
that the Cloitre function, recently introduced by Albert and Kiessling
(\cite{AlKi}, see also \cite{Ki}), allows an exact calculation of 
\textquotedblleft dynamical critical exponents\textquotedblright , which are
seen not to be universal, and, moreover, are proved to be totally
unaccounted for by (rigorous) nonequilibrium mean-field theory.

Section \ref{4} is a conclusion, with a discussion of the main open problems.

Appendix \ref{A} is devoted to an elementary proof \cite{vHBW} of the
existence of a gap in the physical Hamiltonian associated to the forthcoming
\textquotedblleft Ising universality class\textquotedblright\ in Subsection 
\ref{1.2}, see also \cite{KN1}, \cite{KN2}, \cite{KNS}, \cite{NSS}, \cite{DVEPR},
as well as the comments in Subsection \ref{1.4} and Remark \ref{remark:A.1} in
Appendix \ref{A}. 

Appendix \ref{B} provides (we hope) a comprehensive review of the
beautiful work of Newman \cite{Ne1}, which is complementary to ours, in that
it concerns the upper critical dimension.
   
Since we shall base our discussion on the states of infinite systems,
which are the generalization of \eqref{(1.9)} to the whole ``quasilocal''
algebra $\mathcal{A}$, we need to make an interlude in the next subsection
in order to establish some of the fundamental notions. The notions need not be
very extended, because we shall restrict ourselves to the Ising universality
class (the forthcoming generalized Ising model in the case of nearest-neighbor
interactions, the Dyson model in the case of long-range interactions).

\subsection{ Brief mathematical interlude on states of infinite systems}

\label{1.2}

In this section we briefly summarize some of the notions we shall use. The
classic book \cite{Sew1} is a comprehensive, elegant and deep survey. See
also \cite{Hug}, \cite{LvH}, and the treatises \cite{BRo2}, \cite{BRo1}.

A prototype is the generalized Heisenberg Hamiltonian (generalized Ising
model (gIm) if $\delta =0$): 
\begin{equation}
H_{\Lambda }=-2\sum_{x,y\in \Lambda }\left( j(\left\vert x-y\right\vert )(\sigma
_{x}^{3}\sigma _{y}^{3}+\delta H_{XY}(x,y))\right)  \label{(1.2.1)}
\end{equation}
where 
\begin{equation}
H_{XY}(x,y)\equiv \sigma _{x}^{1}\sigma _{y}^{1}+\sigma _{x}^{2}\sigma
_{y}^{2}  \label{(1.2.1a)}
\end{equation}
and 
\begin{equation}
\delta \in (-1,1)  \label{(1.2.1b)}
\end{equation}
and 
\begin{equation}
\sum_{x\in \mathbb{Z}^{\nu }}\left\vert j(x)\right\vert <\infty \text{
\qquad and\qquad\ }j(0)=0  \label{(1.2.2)}
\end{equation}
Above, $\left\vert x\right\vert \equiv \left( \sum_{i=1}^{\nu
}x_{i}^{2}\right) ^{1/2}$, $\nu $ denotes the dimension, and, as asserted in
the introduction, the external field will be set equal to zero in the
present paper. Equation \eqref{(1.2.2)} is the important requirement of 
\textbf{thermodynamic stability}, necessary for the existence of the
thermodynamic limit of the energy per unit volume. Above, $\delta \in (-1,1)$
and we assume 
\begin{equation}
j(\left\vert x\right\vert )\geq 0  \label{(1.2.3)}
\end{equation}
which is appropriate to describe \emph{ferromagnetism}. Above, too, $\sigma
_{x}^{i}$, $i=1,2,3$, are the Pauli matrices at the site $x$, and $
H_{\Lambda }$ acts on the Hilbert space ${\cal H}_{\Lambda }=\otimes
_{x\in \Lambda }\mathbb{C}_{x}^{2}$, and $\vec{S}_{x}$ is short for $\mathbf{
\ 1}\otimes \cdots \otimes \vec{S}_{x}\otimes \cdots \otimes \mathbf{1}$. We
define the algebra of (bounded) operators acting on the Hilbert space $
{\cal H}_{\Lambda }$: ${\cal A}(\Lambda )=B({\cal H}_{\Lambda })$.
These local algebras satisfy: 1.) Causality: $[{\cal A}(B),{\cal A}
(C)]=0$ if $B\cap C=\phi $ and 2.) Isotony: $B\subset C \Rightarrow {\cal
A }(B)\subset {\cal A}(C)$. ${\cal A}_{L}=\cup _{B}{\cal A}(B)$ is
termed the \emph{local} algebra; its closure with respect to the norm, the 
\emph{quasilocal} algebra (observables which are, to arbitrary accuracy,
approximated by observables attached to a \emph{finite} region). The norm is 
$B({\cal H}_{\Lambda })\ni A\longmapsto \left\Vert A\right\Vert
=\sup_{\left\Vert \Psi \right\Vert \leq 1}\left\Vert A\Psi \right\Vert $, $
\Psi \in {\cal H}_{\Lambda }$.

For a finite quantum spin system let $\rho _{\Lambda }$ denote a density
matrix on ${\cal H}_{\Lambda }$, i.e., a non-negative Hermitian matrix of
unit trace on ${\cal H}_{\Lambda }$. We may view $\rho _{\Lambda }$ as a 
\emph{state} $\omega _{\Lambda }$ on $\cal{A}(\Lambda )$ -- a positive,
normed linear functional on ${\cal A}(\Lambda )$: $\omega _{\Lambda }(A)= 
\text{tr}_{{\cal H}_{\Lambda }}(\rho _{\Lambda }A)\text{ for }A \in 
{\cal A}(\Lambda )$ (positive means $\omega _{\Lambda }(A^{\dag }A)\geq 0$,
normed $\omega _{\Lambda }(\mathbf{1})=1$.)

This notion of state generalizes to systems with infinite number of degrees
of freedom $\omega (A)=\lim_{\Lambda \nearrow \infty }\omega _{\Lambda }(A)$,
at first for $A\in {\cal A}_{L}$ and then to ${\cal A}$.

Above, $\lim_{\Lambda \nearrow \infty }$ denotes the thermodynamic limit, in
the simplest case along a sequence of parallelepipeds whose sides tend to
infinity. We shall always consider that $\Lambda =-\Lambda \subset \mathbb{Z}
^{\nu }$.

Considering quantum spin systems on $\mathbb{Z}^{\nu }$, we shall consider
only space-translation-invariant states, i.e., such that 
\begin{equation}
\omega \circ \tau _{x}=\omega \qquad \text{ for all }\qquad x\in \mathbb{Z}
^{\nu }  \label{(1.2.4)}
\end{equation}

As remarked in the previous subsection, an extremal invariant or ergodic
state (see, however, Remark 1.1) is a state which cannot be written as 
a proper convex combination of
two distinct states $\omega _{1}$ and $\omega _{2}$, i.e., the following
does \emph{not} hold: 
\begin{equation}
\omega =a\omega _{1}+(1-a)\omega _{2}\qquad \text{ with }\qquad 0<a<1
\label{(1.2.5)}
\end{equation}
If the above formula is true, it is natural to regard $\omega $ as a mixture
of two pure \textquotedblleft phases\textquotedblright\ $\omega _{1}$ and $
\omega _{2}$, with proportions $a$ and $1-a$, respectively \cite{Wight}.
Alternatively, (see \cite{LvH}, p.403 and references given there), a state $
\omega $ is ergodic with respect to space translations $\tau _{x}$ iff 
\begin{equation}
\lim_{\Lambda \nearrow \infty }\frac{1}{|\Lambda |} \sum_{x\in \Lambda }
\omega (\tau _{x}(A)B) = \omega (A) \omega (B)  \label{(1.2.6)}
\end{equation}

The strong limit $\text{s-lim}$ below is the limit when applied to any
vector in the representation space. The following theorem is
fundamental (\cite{KaRo}, see also \cite{LvH} for an elementary proof):

\begin{theorem}
\label{th:1.2.1} Let $\omega $ be any translation-invariant state. Then the
space average of A 
\begin{equation}
\eta _{\omega }(A)\equiv \slim_{\Lambda \nearrow \infty } \frac{1}{|\Lambda
| }\sum_{x\in \Lambda }\Pi _{\omega }(\tau _{x}(A))  \label{(1.2.8)}
\end{equation}
exists, and, if $\omega $ is ergodic, then 
\begin{equation}
\eta _{\omega }(A)=\omega (A)\mathbf{1}  \label{(1.2.9)}
\end{equation}
\end{theorem}

Equation \eqref{(1.2.9)} corresponds to \textquotedblleft
freezing\textquotedblright\ the observables at infinity to their expectation
values. The space averages $\eta $ defined above correspond to macroscopic
\textquotedblleft pointer positions\textquotedblright , e.g., the mean
magnetization in the Heisenberg model \eqref{(1.2.1)} in the $3$- direction 
\begin{equation}
m_{\Lambda }\equiv \frac{1}{|\Lambda |}\sum_{x\in \Lambda }\sigma _{x}^{3}
\label{(38a)}
\end{equation}
with $A=\sigma ^{3}$. If $\eta _{\omega _{+}}(\sigma ^{3})=a_{+}>0$, and $%
\eta _{\omega _{-}}(\sigma ^{3})= a_{-}<0$, the states $\omega _{\pm }$ are
macroscopically different, i.e., differ from one another by flipping an
infinite number of spins. For a comprehensive discussion, see \cite{Sew1},
Section 2.3.

Finally, a word about \emph{equilibrium states} is necessary (for much more,
see \cite{Hug}). There is an algebra $\tilde{{\cal A}}$ of analytic
elements which is dense in ${\cal A}$, i.e., such that the function $\tau
_{z}(A)$ is an entire function of the complex variable $z$ with values in $
{\cal A}$, and reduces to the time-translation automorphism of ${\cal
A }$ when $z$ is real (Lemmas 4.7 and 4.8 of \cite{Hug}). We say that a
state $\omega $ is a thermal equilibrium state of the system at inverse
temperature $\beta $, or, equivalently, a \emph{KMS state} with respect to $
\tau _{t}$ if 
\begin{equation}
\omega (\tau _{t}(A)B)=\omega (B\tau _{t+i\beta }(A))\qquad \forall B\in 
{\cal A}\text{ \quad and \quad }A\in \tilde{{\cal A}}  \label{(1.2.10)}
\end{equation}
The Gibbs state of a \emph{finite} system at inverse temperature $\beta $ is
given by 
\begin{equation}
\omega _{\beta ,\Lambda }(A)=\text{tr}(\rho _{\Lambda }A)\qquad \text{ with
\quad }\rho _{\Lambda }=\text{ const.}\exp (-\beta H_{\Lambda })
\label{(1.2.11)}
\end{equation}
It is immediate that \eqref{(1.2.11)} satisfies \eqref{(1.2.10)}, but 
\eqref{(1.2.10)} also characterizes \emph{thermal equilibrium states of
infinite systems}, which will be most frequently (but not exclusively, see
Subsection \ref{3.2}) considered in the present paper.

\subsection{Phase transitions in ferromagnetic lattice systems}

\label{1.3}

\subsubsection{Generalities\label{1.3.1}}

For the model given by \eqref{(1.2.1)}, \eqref{(1.2.2)} and \eqref{(1.2.3)}
of ferromagnetic lattice systems, we shall say that the interactions are 
\emph{\ short-range} if the potential $j$ decays exponentially, i.e., 
\begin{equation}
\left\vert j(\left\vert x\right\vert )\right\vert \leq c\exp (-\left\vert
x\right\vert /\Gamma )  \label{(1.3.1)}
\end{equation}
for some $c,\Gamma $ both positive constants. A good approximation of 
\eqref{(1.3.1)} is a \emph{finite-range} interaction, i.e., $j(\left\vert
x\right\vert )=0$ for some $\left\vert x\right\vert \geq e$, $e$ being a
fixed real number, or even a \emph{nearest-neighbor interaction} 
\begin{equation}
j(\left\vert x\right\vert )=0\text{ \qquad unless }\qquad \left\vert
x_{i}\right\vert =1\text{ for all }i=1,\ldots ,\nu  \label{(1.3.2)}
\end{equation}
We shall say that the interactions are \emph{long-range} if they are not
finite-range and there is polynomial decay, i.e., 
\begin{equation}
j(\left\vert x\right\vert )=O(\left\vert x\right\vert ^{-\alpha })\text{
\qquad as\ }\left\vert x\right\vert \longrightarrow \infty \text{ \ for some
\ }\alpha >\nu  \label{(1.3.3)}
\end{equation}
Equation \eqref{(1.3.3)} is due to the stability condition \eqref{(1.2.2)}.
We define the spin-reflection symmetry $\alpha $ by the following equation, valid
for all $x\in \mathbb{Z}^{\nu }$: 
\begin{equation}
\alpha (\sigma _{x}^{1})=\sigma _{x}^{1}\text{ }\qquad \text{together with }
\qquad \alpha (\sigma _{x}^{2})=-\sigma _{x}^{2}\text{ \ and\ }\alpha
(\sigma _{x}^{3})=-\sigma _{x}^{3}  \label{(1.3.4)}
\end{equation}
We have the following theorem:

\begin{theorem}
\label{th:1.3.1}

For the model defined by equations \eqref{(1.2.1)}, \eqref{(1.2.1b)}, 
\eqref{(1.2.2)}, \eqref{(1.2.3)} and \eqref{(1.2.4)}, assume $\nu \geq 2$.
Then, there exists a $T_{c}(\nu )>0$ such that there exist two
\emph{translation-invariant} (extremal) ergodic equilibrium (KMS)
states $\omega _{\pm ,T}^{e}$ for all $T<T_{c}(\nu )$ which are not
invariant under the spin-flip symmetry \eqref{(1.3.4)}, and we say
that a \emph{spontaneous symmetry breakdown} (s.s.b.) occurs. The mean
magnetizations $m_{\pm ,T}$ associated to these states are given by
$\eta _{\omega _{\pm ,T}^{\mathrm{e} }}(\sigma ^{3})\equiv m_{\pm ,T}$
with $m_{+,T}>0$ and $m_{-,T}=-m_{+,T}$. 
\end{theorem}

\noindent \textit{Proof.} The statements of the theorem follow from Theorem
6.2.48 of \cite{BRo2}, together with the remarks before that theorem, and
Theorem \eqref{th:1.2.1}. The assumed translation invariance of the states
is important: see the remarks on the first paragraph of p.338 of \cite{BRo2}
and the references given there.

\hfill $\Box $

\begin{remark}
\label{Remark 1.3.1}

The case of $\delta=0$ corresponds to the Ising model. We shall
consider this boundary case in most of the paper, but follow (\cite{BRo2},
p.320 et seq.) in considering the associated states as states on the whole
algebra $\mathcal{A}$ as before. This choice agrees with the physical
significance of the model in applications, which relies on the fact that it
is an anisotropic limit ($\delta \rightarrow 0$) of the model \eqref{(1.2.1)}.
The dynamics of Subsection 3.2 also takes this choice into account.
When restricting to observables which are functionals of the $\sigma
_{x}^{3} $; $x\in \mathbb{Z}^{\nu }$, however, the correlation inequalities
which are specific to the Ising model, remain, of course, applicable, as
well as the probabilistic language, commonly associated with the
\textquotedblleft classical version\textquotedblright .

Theorem~\ref{th:1.3.1} substantiates considering the model defined by
equations \eqref{(1.2.1)} as belonging to \emph{the Ising universality class}. An
additional reason for this nomenclature is the property of \emph{Ising
domination} proved in Appendix \ref{A}.

Although we keep the dimension $\nu$ general, for clarity, we shall only
consider , when dealing with short-range interactions, the 
(physically relevant) cases $\nu=2$ and $\nu=3$ in correspondence
with Theorem~\ref{th:1.3.1}. An upper bound to the upper critical dimension is $\nu_{c}=4$
\cite{ADC}. For high dimensions $\eta=0$, and the systems rescales to a
``massless'' Gaussian, which is quasi-free, but has connected correlations decaying
as $r^{2-\nu}$. The very recent paper by Duminil-Copin and Panis \cite{DCP} proves
that the critical exponent $\eta=0$ for $\nu$ at least 4, and gives improved rigorous
bounds on $\eta$ for $\nu=3$.

\end{remark}

\subsubsection{A criterion of criticality: Criterion A}

\label{1.3.2}

The critical state $\omega _{T_{c}}$ is the unique, spin-flip invariant --
and hence ergodic -- state at $T=T_{c}$. By Example 4.3.24 of \cite{BRo1},
together with the states $\omega _{\pm ,T}^{\mathrm{e}}$ in
Theorem~\ref{th:1.3.1}, it also satisfies the property of clustering
in space:  
\begin{equation}
\lim_{\left\vert x\right\vert \rightarrow \infty }\omega _{T,C}^{\mathrm{e}
}(\sigma ^{3}\tau _{x}(\sigma ^{3}))=0  \label{(1.3.7)}
\end{equation}
with the notation \eqref{(1.10)}. The above equation \eqref{(1.3.7)} is
therefore valid for all $0<T\leq T_{c}$. It is now natural to inquire into
the \emph{rate} of clustering. It has recently been proved in a remarkable
paper \cite{DCGR} that 
\begin{equation}
0\leq \omega _{T,C}^{\mathrm{e}}(\sigma ^{3}\tau _{x}(\sigma ^{3}))\leq \exp
(-c_{T}|x|)\qquad \forall x\in \mathbb{Z}^{\nu }  \label{(1.3.12a)}
\end{equation}
with 
\begin{equation}
c_{T}>0\qquad \forall \quad 0<T<T_{c}\ .  \label{(1.3.12b)}
\end{equation}
The quantity $c_{T}$ in \eqref{(1.3.12a)} has the interpretation of
\textquotedblleft inverse correlation length\textquotedblright , and, since
it is also strictly positive for $T>T_{c}$ (see \cite{DCGR} and references
given there), it may tend to zero \emph{only} at the critical point. This
possibility, and its implications for the phenomenological theory of scaling
and critical fluctuations (see \cite{Sew1}, 5.4.1, and references given
there), will be one of our major concerns in the present paper: it has been
analysed by Sewell \cite{Sew1}, p. 122, section 5.4), who suggested the
following argument.

Let 
\begin{equation}
\bar{\omega}_{T}\equiv \frac{1}{2}(\omega _{+,T}^{\mathrm{e}}+\omega
_{-,T}^{ \mathrm{e}})  \label{(1.3.13a)}
\end{equation}
The state $\bar{\omega}_{T}$ is spin-flip and translation invariant, and, by
Theorem \ref{th:1.3.1} it follows immediately that 
\begin{equation}
\lim_{\Lambda \nearrow \mathbb{Z}^{\nu }}\frac{1}{|\Lambda |^{2}}\bar{\omega}
_{T}\left( \left( \sum\nolimits_{x\in \Lambda }\sigma _{x}^{3}\right)
^{2}\right) =\left\vert m_{\pm ,T}\right\vert ^{2}>0  \label{(1.3.13)}
\end{equation}
as long as $0<T<T_{c}$, and, hence, 
\begin{equation}
\lim_{\Lambda \nearrow \mathbb{Z}^{\nu }}\frac{1}{\left\vert \Lambda
\right\vert }\sum_{x,y\in \Lambda }\bar{\omega}_{T}(\sigma _{x}^{3}\sigma
_{y}^{3})=\infty  \label{(1.3.14)}
\end{equation}
for $T<T_{c}$. By translation invariance of $\bar{\rho}$, \eqref{(1.3.14)}
implies that 
\begin{equation}
\sum_{y\in \mathbb{Z}^{\nu }}\bar{\omega}_{T}(\sigma _{0}^{3}\sigma
_{y}^{3})=\infty \text{ for }T<T_{c}  \label{(1.3.15)}
\end{equation}
Since $\bar{\omega}_{T}(\sigma _{x}^{3})=0$, $\forall x\in \mathbb{Z}^{\nu }$,
it is natural to assume that \eqref{(1.3.15)} holds, by continuity, up to
the critical point, the critical state being unique and thus also invariant
under the spin-flip symmetry: 
\begin{equation}
\sum_{x\in \mathbb{Z}^{\nu }}\omega _{T_{c}}^{\mathrm{e}}(\sigma ^{3}\tau
_{x}(\sigma ^{3}))=\infty  \label{(1.3.16)}
\end{equation}
i.e., the \textquotedblleft clustering is not summable at the critical
point\textquotedblright . Since $\omega _{T_{c}}$, being unique, is also
ergodic, equations \eqref{(1.3.12a)}, \eqref{(1.3.12b)} and \eqref{(1.3.16)}
suggest the following criterion:

\emph{Criterion A} The critical state is singled out, among all ergodic
states for $T \le T_{c}$, by the condition that its \emph{connected
two-point function is not summable}.

Criterion A is not new: it agrees with the ``renormalization group and
probability theory'' approach of several researchers, see the review by
Jona-lasinio \cite{JL} and references given there.

\begin{remark}
\label{Remark 1.3.2}
The uniqueness of critical state is related with another important issue
regarding the continuity of spontaneous magnetization for ferromagnetic
Ising model, established recently for $d=3$. See \cite{ADCV} and references
therein.
\end{remark}

\subsubsection{SLRO, LLRO and their connection with Criterion A}

\label{1.3.3}

We now come back to the notions of normalized fluctuation operator 
\eqref{(1.18)}, \eqref{(1.19)}, \eqref{(1.21)} and \eqref{(1.22)}, and the
corresponding characteristic functions (ch.f.) $\Phi _{\Lambda ,T}(t)$ and $
\tilde{\Phi}_{\Lambda ,T}(t)$, given by \eqref{(1.23)} and \eqref{(1.24)},
respectively. The two different limits $\Phi _{T}$ and $\tilde{\Phi}_{T}$,
given by \eqref{(1.25)} and \eqref{(1.26)} (when they exist) are known as
short-long-range-order (SLRO) and long-long-range order (LLRO), respectively 
\cite{SML}. They need not coincide - the fact that the normalized
fluctuation operators are not uniformly bounded in norm leaves this as an a
priori possibility, but explicit examples have in fact been given in
mean-field theories, by Ellis, Newman and Rosen \cite{ENR}, see also the
explicit example in \cite{MCJL1}, which has been abstracted from \cite{ENR}.
We conjecture, however, that LLRO and SLRO agree for short-range
interactions. This conjecture is relevant to Theorem \ref{th:2.3.1}.

\begin{remark}
\label{Remark 1.4.2} In order to analyse the transition from $0<T<T_{c}$ to $
T=T_{c}$, which is implicit in Criterion A, it is necessary (see \cite{DCGR})
to consider one of the ergodic states $\omega _{\pm ,T}^{e}$ (for which
the mean magnetization $m\neq 0$), and $\omega _{T_{c}}$, for $T=T_{c}$. If
the alternative $\Phi _{\Lambda ,T}$, as $\Lambda \nearrow \mathbb{Z}^{\nu }$,
is considered (equivalent to LLRO \cite{SML}), boundary conditions $\pm $
must be taken at $\Lambda $ for $T<T_{c}$, and free or periodic b.c. at $
T=T_{c}$, and a fixed choice, which imposes itself to define the procedure
in precise mathematical terms is not adequate to cover both cases. Since
different ergodic states of infinite systems are typically associated to
performing the thermodynamic limit with different choices of b.c., as
indicated above, we see that, in addition to the arguments employed in
Subsection \ref{1.1}, the choice of regarding the states of infinite systems
as the basic objects of the theory is the only natural one.
\end{remark}

\subsubsection{ The connected Wightman functions}
\label{1.3.4}

We now come back to \eqref{(1.23)} and \eqref{(1.25)} and define

\begin{equation}
\left. \frac{d^{r}}{dt^{r}}\log \Phi _{A,\Lambda }(t)\right\vert _{t=0}=\sum 
_{\substack{ x_{1},\ldots ,x_{r}:  \\ x_{i}\in \Lambda ,~i=1,...,r}}
W_{\omega _{T},C}^{r}\left( x_{1},\ldots ,x_{r}\right)  \label{(2.1.1a)}
\end{equation}
where the $W_{\omega _{T},C}^{r}(x_{1},\cdots ,x_{r})$ are the connected (or
truncated) Wightman (or correlation) $r$-point functions associated to the
state $\omega _{T}$: 
\begin{equation}
W_{\omega ,C}^{1}\left( x_{1}\right) \equiv \omega \left( A_{x_{1}}\right)
\label{(2.1.1b)}
\end{equation}
\begin{equation}
W_{\omega ,C}^{2}\left( x_{1},x_{2}\right) \equiv \omega \left(
A_{x_{1}}A_{x_{2}}\right) -\omega \left( A_{x_{1}}\right) \omega \left(
A_{x_{2}}\right)  \label{(2.1.1c)}
\end{equation}
and, generally, by recursion,
\begin{equation}
W_{\omega ,C}^{r}\left( x_{1},\ldots ,x_{r}\right) \equiv \omega \left(
A_{x_{1}}\cdots A_{x_{r}}\right) -\sum_{\pi _{1},\ldots ,\pi
_{k}}\prod_{j=1}^{k}W_{\omega ,C}^{j}
\left( x_{i_{1}} \ldots x_{i_{j}} \right)  \label{(2.1.1d)}
\end{equation}
at $\omega =\omega _{T}$, where the sum in \eqref{(2.1.1d)} runs over all
non-trivial (i.e., $k>1$) partitions of $x_{1},\ldots ,x_{r}$ into subsets $
\pi _{1},\ldots ,\pi _{k}$. We replaced the notation $\Phi _{\Lambda }(t)$
in \eqref{(1.23)} by $\Phi _{A,\Lambda }(t)$, specifying the observable $A$
which is arbitrary, but in \eqref{(1.23)} was $A=\sigma ^{3}$. This choice
will be made in this section, because we shall consider the Ising model ($
\delta =0$ in \eqref{(1.2.1)}), but keeping it general at this stage serve
the purpose of clarity. Notice that \eqref{(2.1.1c)} is consistent with 
\eqref{(1.10)}, in this notation, with $A=B$.

If $A=\sigma ^{3}$, and $\delta =0$ in \eqref{(1.2.1)}, $W_{\omega
_{T},C}^{r}$ where $\omega _{T}$ is any equilibrium state at temperature $T$
of the Ising model, two very important inequalities will concern us: 
\begin{equation}
W_{\omega ,C}^{2}(x_{1},x_{2})\geq 0  \label{(2.1.2)}
\end{equation}
and 
\begin{equation}
W_{\omega ,C}^{4}(x_{1},\ldots ,x_{4})\leq 0  \label{(2.1.3)}
\end{equation}
equation \eqref{(2.1.2)} is Griffiths' second inequality \cite{Gri},
equation \eqref{(2.1.3)} is Lebowitz' inequality \cite{Leb}. See also \cite{GJ}.

Shlosman \cite{Shl} proved the general statement about the signs for higher-order Ursell functions.

\subsubsection{Interacting and quasi-free theories}
\label{1.3.5}

Some of the most difficult problems in the theory of the critical point
concern the issue of triviality and the connections with quantum field
theory. In the lattice-gas picture, model \eqref{(1.2.1)} has an interesting
interpretation (see \cite{BRo2}, p.425 et seq): the XY term (which
multiplies $\delta$ in \eqref{(1.2.1)}) is a discrete form of the Laplacean
and is interpretable as a quasi-free or non-interacting lattice gas, while
the Ising term corresponds to a potential energy arising from a two-body
interaction between neighboring particles.

In the lattice-gas picture, with the spin-raising and lowering
operators (\cite{BRo2}, p. 425) (in dimension $\nu =1$)  
\begin{equation}
a_{x}^{\ast }=\frac{\sigma _{x}^{1}+i\sigma _{x}^{2}}{2}  \label{(1.3.5a)}
\end{equation}
and 
\begin{equation}
a_{x}=\frac{\sigma _{x}^{1}-i\sigma _{x}^{2}}{2}  \label{(1.3.5b)}
\end{equation}
there arises a Hamiltonian term 
\begin{equation*}
H=-2J\sum_{x=-n}^{n}a_{x}^{\ast }a_{x+1}^{\ast }a_{x+1}a_{x}
\end{equation*}%
which corresponds to a potential energy from a two-body interaction between
neighboring particles, in contrast to $H_{XY}$ in \eqref{(1.2.1a)}, which is
quadratic in the creation and annihilation operators $a_{x}$, $a_{x}^{\ast }$,
$x\in \mathbb{Z}$, see (\cite{BRo2}, p.426). This latter case is an
example of a quasi-free system, the former of an interacting system. We
define precisely:

\begin{definition}
\label{Definition 1.3.5}
A system defined by the connected Wightman functions $W_{\omega
_{T},C}^{r}\left( x_{1},\ldots ,x_{r}\right) $, given by \eqref{(2.1.1b)}, 
\eqref{(2.1.1c)}, and \eqref{(2.1.1d)} is called quasi-free if $W_{\omega
_{T},C}^{r}\left( x_{1},\ldots ,x_{r}\right) \equiv 0$ for $r>2$, otherwise
it is called an interacting system.
\end{definition}

The thermal Wightman functions of the Ising model (\eqref{(1.2.1)} with $
\delta = 0$) are expected to define an interacting system in the sense of
Definition ~\ref{Definition 1.3.5}, except if $T=0$. In this case, the
ergodic states are (infinite) product states, see also Appendix A.
Otherwise, the connected four-point
Wightman function is not expected to be identically zero. This is because this 
would entail that all higher connected functions vanish, and hence already for finite $|\Lambda|$
the characteristic function would be a Gaussian, a behavior not expected for the Ising model
which may be viewed as a system of interacting ``fermions'' in the sense of (\cite{BRo2}, p. 426).

\begin{remark}
\label{Remark 1.3.5}

The importance of the interacting nature of the Ising system in our case is
that the associated theory of phase transitions is physically a sound one,
i.e., there are no non-physical features, as, e.g., in the phase transition
in the free Bose gas, due to infinite compressibility, see, for instance, 
\cite{Dor}, pp. 200-209. It is also related to the previously mentioned
success of the Ising model in the description of real experiments. This
property concerns the equilibrium states; in this paper, we shall be
concerned, however, with the quasi-free or interacting nature of the
macroscopic states, as explained in the next subsection.
\end{remark}

\subsection{Macroscopic phenomena}

\label{1.4}

One of the deepest problems in physics is how macroscopic phenomena can be
interpreted from the laws and structures of microphysics \cite{Sew2}. One
feature of this problem is to account for the observed macroscopic
irreversibility, while the equations of microphysics are time-reversal
invariant (\cite{Wre1}, \cite{Wre2}). Another one is the construction of a
sensible macroscopic measuring apparatus in the theory of quantum mechanical
measurement (see \cite{Wre3}: there a sequence of pure states for a system
in finite volume tends to a mixed state in the thermodynamic limit.

Two observables reflecting macroscopic behavior are \emph{intensive
observables}, such as the mean magnetization \eqref{(38a)}, and the \emph{\
fluctuation observables}, such as \eqref{(1.18)} and \eqref{(1.21)}.

A state which is the limit of 
\begin{equation*}
\left( \left( a\Psi _{1,\Lambda }+b\Psi _{2,\Lambda }\right) , (.)\left( a\Psi
_{1,\Lambda }+b\Psi _{2,\Lambda }\right) \right)
\end{equation*}
where $a$ and $b$ are complex numbers, such that 
\begin{equation*}
|a|^{2}+|b|^{2}=1
\end{equation*}
becomes, in the limit $\Lambda \nearrow \mathbb{Z}^{\nu }$, under certain
conditions (see \cite{Wre3} and references given there) a state of the form 
\begin{equation*}
|a|^{2}(\Psi _{1},(.)\Psi _{1})+|b|^{2}(\Psi _{2},(.)\Psi _{2})
\end{equation*}
That is, the relative phase of $b$ and $a$ in $a\Psi _{1,\Lambda }+b\Psi
_{2,\Lambda }$ disappears in the (proper) macroscopic limit, leading to the
mixed state \eqref{(1.2.5)}.

In the theory of phase transitions (see Subsection \ref{1.3.1},
Theorem \ref{th:1.3.1} shows that \eqref{(38a)}, the mean
magnetization is an order parameter which distinguishes the two
microscopic states $\omega _{\pm ,T}^{\mathrm{e}}$ (phases) for $0<T<T_{c}$, and 
\begin{equation}
\omega _{T}\equiv a\omega _{+,T}^{\mathrm{e}}+(1-a)\omega _{-,T}^{\mathrm{e}
} \text{ \qquad with \qquad }0<a<1  \label{(1.4.1)}
\end{equation}
represents a \emph{classical} mixture. This remark applies to both the
magnetic and the lattice fluid pictures, of course, but, in the latter case,
when the temperature $T\nearrow T_{c}$, the resulting mixture displays an
outstanding classical (macroscopic) effect - the critical opalescence -
discussed in Subsection \ref{1.1}. In the former case, it is known that
domains of a magnet are, in nature, found only in a \emph{definite}
direction. This constitutes one of the rare examples
for which the existence of so-called \emph{superselection sectors} \cite{Wight1} originating from
interactions with the environment, which are observed in nature, may be proved: a rigorous model 
of this situation may be
constructed precisely around \eqref{(1.2.1)} following the model suggested
by Narnhofer and Thirring \cite{NT}, see \cite{Wre2}, Section 4.2, and
references given there, as well as \cite{NT} and \cite{NW}. An alternative
approach to this problem was suggested in \cite{MR1}, in which it was shown
that every generic many-body system contains within the class of microscopic
quantum observables a subalgebra of macro-observables, the spectrum of which
comprises the macroscopic properties of the many-body system (see also
\cite{GL}). A third, independent, approach to show the existence of
domain walls was initiated in \cite{ASW} and further considerably
extended by Koma, Nachtergaele, 
Spitzer and Starr (\cite{KN1}, \cite{KN2}, \cite{KNS}, \cite{NSS}), see also 
Remark \ref{remark:A.1} in Appendix \ref{A}.

Concerning the fluctuations \eqref{(1.18)}, \eqref{(1.21)}, it is remarkable
that they retain a quantum character in spite of the thermodynamic limit: in
fact, they are related to (Bosonic) spin waves (\cite{Ver}, \cite{MR}).
consider the model \eqref{(1.2.1)} and, coming back to \eqref{(1.3.5a)}, 
\eqref{(1.3.5b)}, define 
\begin{equation}
a_{\Lambda }\equiv \frac{1}{\sqrt{\left\vert \Lambda \right\vert }}
\sum_{x\in \Lambda }a_{x}  \label{(1.4.2a)}
\end{equation}
and 
\begin{equation}
a_{\Lambda }^{\ast }\equiv \frac{1}{\sqrt{\left\vert \Lambda \right\vert }}
\sum_{x\in \Lambda }a_{x}^{\ast }  \label{(1.4.2b)}
\end{equation}
We have, by Theorem \ref{th:1.2.1} and Theorem \ref{th:1.3.1}, 
\begin{equation}
\lim_{\Lambda \nearrow \mathbb{Z}^{\nu }}\omega _{\pm ,T}^{\mathrm{e}
}([a_{\Lambda },a_{\Lambda }^{\ast }])=m_{\pm ,T}  \label{(1.4.3)}
\end{equation}
since $[a_{\Lambda },a_{\Lambda }^{\ast }]=\left. \sum_{x\in \Lambda }\sigma
_{x}^{3}\right/ \left\vert \Lambda \right\vert $. Thus, for $0<T<T_{c}$, the
quantities 
\begin{equation}
\left( \left\vert m_{\pm }\right\vert \left\vert \Lambda \right\vert \right)
^{-1/2}\sum_{x\in \Lambda }\exp (ik\cdot x)a_{x}  \label{(1.4.4a)}
\end{equation}
and 
\begin{equation}
\left( \left\vert m_{\pm }\right\vert \left\vert \Lambda \right\vert \right)
^{-1/2}\sum_{x\in \Lambda }\exp (-ik\cdot x)a_{x}^{\ast }  \label{(1.4.4b)}
\end{equation}
with $-\pi \leq k_{i}<\pi ,i=1,\ldots ,\nu $, are expected to define Boson
(spin-wave) operators $a_{B}(k)$, $a_{B}(k)^{\ast }$ in the limit $\Lambda
\nearrow \mathbb{Z}^{\nu }$, in the representation spaces of the states $%
\omega _{+,T}^{\mathrm{e}}$, for $0<T<T_{c}$, with $a_{B}^{\ast }$ replacing 
$a_{B}$ in the state $\omega _{-,T}^{\mathrm{e}}$. Equations \eqref{(1.4.4a)}%
, \eqref{(1.4.4b)} exhibit the basic role of fluctuation operators as the
links between quantum mechanics and macroscopic systems. This fact has been
specially emphasized in \cite{Ver} and \cite{MR}, and, previously, by Hepp
and Lieb in their seminal paper on phase transitions out of thermal
equilibrium \cite{HL}. There, they define precisely the limiting state $%
\omega _{T}^{F}$ (see also \cite{MR}):

\begin{definition}
\label{Definition 1.4.1}

Assume that the limits 
\begin{equation}
\lim_{\Lambda \nearrow \mathbb{Z}^{\nu }}\omega _{T}\left( F_{\Lambda
,T}^{n}\right) \qquad \text{ with \qquad }n=1,2,3,\cdots  \label{(1.4.5)}
\end{equation}
exist. Above, if $\delta =0$ in \eqref{(1.2.1)}, $F_{\Lambda ,T}$ are
defined by \eqref{(1.21)} if SLRO is chosen, otherwise \eqref{(1.18)} must
be chosen, with $\omega _{T}$ replaced by $\omega _{\Lambda ,T}$ in %
\eqref{(1.4.5)}. We assume here the obvious extension of the definition to
the full model \eqref{(1.2.1)}, with three fluctuation operators $F_{\Lambda
,T}^{i},i=1,2,3$ defined in analogy to \eqref{(1.4.5)}: 
\begin{equation}
F_{\Lambda ,T}^{i}\equiv \frac{1}{\sqrt{\left\vert \Lambda \right\vert }}
\sum_{x\in \Lambda }\left( \sigma _{x}^{i}-\omega _{T}\left( \sigma
_{0}^{i}\right) \right) \qquad \text{ with \quad }i=1,2,3  \label{(1.4.6)}
\end{equation}
Then, the state $\omega _{T}^{F}$, which we call the \emph{fluctuation state},
is the state defined by the GNS construction (see \cite{BRo1}, p.153) on
the polynomial algebra $\mathcal{P}$ \cite{HL}, generated by new limiting
objects, the fluctuation operators $F_{T}$, by 
\begin{equation}
\omega _{T}^{F}\left( \left( F_{T}^{i}\right) ^{n}\right) =\lim_{\Lambda
\nearrow \mathbb{Z}^{\nu }}\omega _{T}\left( \left( F_{\Lambda
,T}^{i}\right) ^{n}\right) \qquad \text{ with \quad }n=1,2,\ldots \quad 
\text{and}\quad i=1,2\ .  \label{(1.4.7)}
\end{equation}
\end{definition}

Note that the original equilibrium state $\omega_{T}$ is a state on a
different algebra of observables, the quasi-local algebra $\mathcal{A}$.

We know of only two concrete constructions of fluctuation states $%
\omega_{T}^{F}$: Hepp and Lieb's \cite{HL} and Wehrl's \cite{Wehrl}.

The spin-wave operators \eqref{(1.4.4a)}, \eqref{(1.4.4b)} provide an
accurate description of the low-temperature magnetization of ferromagnets 
\cite{Mat}. More recently, the temperature dependence of the junction
critical Josephson current in the strong-coupling regime was obtained by
using a quantum fluctuation description of charge qubits \cite{BCFNV}, whose
basic elements were collective quasi-spin fluctuation operators, which
generate the Heisenberg algebra on the circle in a model of the
strong-coupling BCS type (see, e.g., \cite{Dor}, Chapter 35).

The convergence of \eqref{(1.4.4a)}, \eqref{(1.4.4b)} to Boson spin-wave
operators $a_{B}(k)$, $a_{B}^{\ast }(k)$ in the representation space of the
states $\omega _{\pm ,T}^{\mathrm{e}}$ is true under the assumption 
\eqref{(1.4.5)} of Definition \ref{Definition 1.4.1} for, then, the limiting
objects $F_{T}^{1,2}$ must satisfy \eqref{(1.4.3)}. For $0\leq T<T_{c}$, the
operators $F_{\Lambda ,T}^{3}$ are defined as in \eqref{(1.4.6)}, with the
normalization by $1/\sqrt{\left\vert \Lambda \right\vert }$; as such, they
define a Lie algebra, the \emph{Lie algebra of fluctuations} (\cite{Ver}, 
\cite{VZ}), whose importance was also emphasized in \cite{HL}; in the
present case, it is the algebra $su(2)$, the Lie algebra of the group $SU(2)$.
If we assume that, in addition to \eqref{(1.4.3)} and the fact that $
m_{\pm ,T}=0$ at $T=T_{c}$, 
\begin{equation}
\lim_{\Lambda \nearrow \mathbb{Z}^{\nu }}\omega _{\pm ,T}^{\mathrm{e}
}(\left. \sigma _{x}^{i}\right/ |\Lambda |)=0\qquad \text{ for \quad }i=1,2\,
\label{(1.4.8)}
\end{equation}
it follows that the thermodynamic limit induces a \emph{contraction of Lie
algebra representations} (see \cite{CW1}, \cite{CW2}), viz. that of 
\begin{equation}
su(2)\rightarrow h(2)  \label{(1.4.9)}
\end{equation}
(\cite{CW2}, III.1.2), $h(2)$ denoting the two-dimensional Heisenberg (CCR)
algebra. Since $m_{\pm ,T}=0$ at $T=T_{c}$, and assuming anomalous
fluctuations \eqref{(1.20)} in the 3-direction, the passage to $T\nearrow
T_{c}$ leads to an additional contraction 
\begin{equation}
h(2)\rightarrow r^{3}  \label{(1.4.10)}
\end{equation}
(\cite{CW2}, III.1.3), where $r^{3}$ denotes the three-dimensional abelian
Lie algebra, isomorphic to the Lie algebra of $\mathbb{R}^{3}$. If, in 
\eqref{(1.4.6)} the factors $1/\sqrt{\left\vert \Lambda \right\vert }$ are
both replaced with $1/\left\vert \Lambda \right\vert ^{\alpha }$, with $
\alpha >1/2$, i.e., the fluctuations of $\sigma ^{1,2}$ are also anomalous, 
\eqref{(1.4.10)} is also true a fortiori: see \cite{Sew1}, Appendix \ref{B},
p.166, for a general theorem in this connection. Equations \eqref{(1.4.9)}
and \eqref{(1.4.10)} together imply that the critical fluctuation state $
\omega _{T}^{F}$ of the ferromagnets in the Ising universality class 
\eqref{(1.2.1)} is characterized, from the point of view of the algebra of
fluctuations, as the unique state for which the \emph{Lie algebra of
fluctuations} is \emph{classical}. This explains why critical behavior
in quantum models may be described by classical Ising models.

We refer also to the references on the extensive work on the quantum
central limit theorem given in the book by Verbeure \cite{Ver}.

\begin{remark}
\label{Remark 1.4.1}

A final, most important, aspect of the relation between the states $%
\omega_{T}^{F}$ and $\omega_{T}$ is the possible occurrence of a \emph{basic
structural difference} between them, taking into account, of course, that
they are states on different algebras of observables, $\mathcal{P}$ and $%
\mathcal{A}$, as previously remarked. One possibility, proved under certain
natural assumptions in the forthcoming Theorem \ref{th:2.3.1}, is that $%
\omega_{T}$ is an interacting state, while $\omega_{T}^{F}$ is a quasi-free
state. We shall now try to provide some intuition, why this may be expected
to occur.

The bosonic quasi-spin operators \eqref{(1.4.4a)}, \eqref{(1.4.4b)}
(spin-waves, with $\left\vert m_{\pm }\right\vert =1$) have been studied in
great detail by Wehrl in the BCS model \cite{Wehrl}. In a certain limit (the
Trotter limit, also used in \cite{CW1}, \cite{CW2}), their \emph{dynamics}
is shown to satisfy the KMS condition \eqref{(1.2.10)} only if the value $%
k=0 $ is excluded in \eqref{(1.4.4a)}, \eqref{(1.4.4b)}. The zero-momentum
fluctuation operators are invariant under all permutations of the elements
of the set $\{x\in \Lambda \}$, as is its Hamiltonian (which is the BCS
mean-field Hamiltonian), and, comparing this situation with ordinary
many-body systems, one sees that the operators $a_{\Lambda }$, $a_{\Lambda
}^{\ast }$, in \eqref{(1.4.2a)}, \eqref{(1.4.2b)}, are the analogues of the
center of mass variables, and thus behave \emph{freely}. Consequently, they
do not \textquotedblleft thermalize\textquotedblright , and this happens for
all temperatures. If we believe that these considerations apply,
qualitatively speaking, to general models, we are led to conjecture that,
for all temperatures (including $T_{c}$), the fluctuation state is
quasi-free.

Furthermore, if we assume that the fluctuation state $\omega_{T}^{F}$ is a
rigorous version of a ``re\-nor\-ma\-li\-za\-tion-group (RG)
transformation'' (in real space), the last assertion in the previous
paragraph implies that RG transformations are \emph{singular}, in the sense
that $\omega_{T}^{F}$ and $\omega_{T}$ display a basic structural
difference. Moreover, not being an equilibrium (KMS) state for \emph{any}
temperature (since any such state is an interacting state), it need not be
locally Gibbs (\eqref{(1.2.11)}). The remarks in this paragraph provide a
(possible) connection to the RG peculiarities pointed out by Griffiths and
Pearce \cite{GP}, see also van Enter's comprehensive review \cite{vE} and
the huge literature cited there, and Sewell's pedagogic account in (\cite%
{Sew1}, 5.4.4).
\end{remark}

\subsection{Ferromagnetic Ising systems in finite volume and infinitely
divisible characteristic functions}

\label{1.5}

In this section we study the characteristic function $\tilde{\Phi}_{\Lambda
,T}(t)$, given by \eqref{(1.24)}, for finite volume. Our basic reference is
the important paper of De Coninck and Gottal \cite{dCG}. Recall that a
characteristic function (ch.f.) $f(t)$ is a Fourier-Stieltjes transform of a
distribution function $dF$, defined for all $t\in \mathbb{R}$ by (\cite%
{Chung}, Chap. 6.1, p. 142) 
\begin{equation}
f(t)=\int \exp \left( itx\right) dF(x)\ .  \label{(1.5.1)}
\end{equation}
A ch.f. is called \emph{infinitely divisible} iff, for all integers $n\geq 1$%
, there exist ch.f. $f_{n}$ such that 
\begin{equation}
f(t)=\left( f_{n}(t)\right) ^{n}  \label{(1.5.2)}
\end{equation}
(see \cite{Chung}, 7.6, or \cite{Lu}, 108, 109). Instead of $\tilde{\Phi}%
_{\Lambda ,T}(t)$, we consider 
\begin{equation}
\Phi _{T}^{\Lambda }(t)\equiv \exp \left( it\sum_{x\in \Lambda }\omega
_{\Lambda ,T}\left( \sigma _{x}^{3}\right) \right) \tilde{\Phi}_{\Lambda
,T}(t) \, ,
\end{equation}
or 
\begin{equation}
\Phi _{T}^{\Lambda }(t)=\omega _{\Lambda ,T}\left( \exp \left(
it\sum\nolimits_{x\in \Lambda }\sigma _{x}^{3}\right) \right) \ .
\label{(1.5.3)}
\end{equation}
For finite $\Lambda $, and real $t$, we have (\cite{dCG}, Lemma 1): 
\begin{equation}
\Phi _{\Lambda }^{T}(t)=\prod_{j=1}^{\infty }\left( 1+\frac{t^{2}}{t_{j}^{2}}
\right)  \label{(1.5.4)}
\end{equation}
where $\left\{ t_{j}\right\} _{j={1}}^{\infty }$ is an \emph{infinite} set
of real numbers such that 
\begin{equation}
0<t_{1}\leq t_{2}\leq \cdots  \label{(1.5.5)}
\end{equation}
Notice that \eqref{(1.5.4)} is invariant by inclusion of the variance in %
\eqref{(1.18)}, since $t\rightarrow t/\sqrt{V_{\Lambda ,T}}$ together with $%
t_{j}\rightarrow t_{j}/\sqrt{V_{\Lambda ,T}}$. The following arguments are
due to De Coninck and Gottal in \cite{dCG}. By \eqref{(1.5.4)}, the
reciprocal of $\Phi _{\Lambda }(t)$, 
\begin{equation}
\frac{1}{\Phi _{T}^{\Lambda }(t)}=\prod_{j=1}^{\infty }\left( 1+\frac{t^{2}}{
t_{j}^{2}}\right) ^{-1}  \label{(1.5.6)}
\end{equation}
De Coninck and Gottal noticed that the quantity $\left(
1+t^{2}/t_{j}^{2}\right) ^{-1}$ is the ch.f. of the Laplace probability
distribution, viz. $\exp (-\left\vert x\right\vert )$, over $x\in \mathbb{R}$%
, which is infinitely divisible (\cite{Lu}, pp. 108, 109). Since the limit
of a sequence of infinitely divisible ch.f. is infinitely divisible (\cite%
{Lu}, Theorem 5.3.3, p. 110), $1/\Phi _{\Lambda }^{T}(t)$ is infinitely
divisible. This result is the essential ingredient of Theorem 2 of \cite{dCG}%
, whose consequences will be mentioned shortly. By \eqref{(1.5.3)} and %
\eqref{(2.1.1a)}, we have, connecting our notation with that ($U_{r,\Lambda
}(0)$) of \cite{dCG}, 
\begin{eqnarray*}
\left( \frac{d^{r}}{dt^{r}}\log \Phi _{\Lambda }^{T}\right) (0)
&=&-U_{r,\Lambda }(0) \\
&\equiv &i^{r}\sum_{\substack{ x_{1},\ldots ,x_{r};  \\ x_{i}\in \Lambda
,~i=1,...,r}}W_{\omega _{\Lambda ,T},C}^{r}\left( x_{1},\ldots ,x_{r}\right) 
\end{eqnarray*}
Correspondingly, 
\begin{equation}
U_{r,\Lambda }(0)=-i^{r}\sum_{x_{1},\ldots ,x_{r};x_{j}\in \Lambda
}W_{\omega _{\Lambda ,T},C}^{r}\left( x_{1},\ldots ,x_{r}\right)
\label{(1.5.8)}
\end{equation}
Notice the difference of sign with respect to the previous equation, which
corresponds to taking the reciprocal in \eqref{(1.5.6)}. By Theorem 2 of 
\cite{dCG}, 
\begin{equation}
U_{2r+2,\Lambda }(0)=(-1)^{r}\int_{-\infty }^{\infty }x^{2r}dK_{\Lambda }(x)
\label{(1.5.9)}
\end{equation}
where 
\begin{equation}
K_{\Lambda }(+\infty )=2\sum_{j=1}^{\infty }\frac{1}{t_{j}^{2}}
\label{(1.5.10)}
\end{equation}
We have thus the important corollary

\begin{proposition}
\label{prop:1.5.1} $\left\vert U_{2r+2,\Lambda }(0)\right\vert \neq 0$ for
all $r=0,1,2,\cdots $
\end{proposition}

\noindent \textit{Proof.} By \eqref{(1.5.9)}, $U_{2r+2,\Lambda} \equiv 0$
iff $dK_{\Lambda}(x) = \lambda \delta(x)$ for some $\lambda > 0$, and $%
\delta $ the delta measure at zero, but $K_{\Lambda}(\infty) \neq 0$ by %
\eqref{(1.5.10)}.

\hfill $\Box $

A second result which will be useful for the forthcoming Theorem \ref%
{th:2.3.1} is:

\begin{proposition}
\label{prop:1.5.2} Assume that: 
\begin{equation}
\text{ there exists\quad\ }\lim_{\Lambda \nearrow \mathbb{Z}^{\nu
}}U_{2r,\Lambda }(0)\quad \text{ for all \quad }r=1,2,3,\cdots
\label{(1.5.11)}
\end{equation}
Then, if 
\begin{equation}
\lim_{\Lambda \nearrow \mathbb{Z}^{\nu }}U_{2r,\Lambda }(0)\equiv 0
\label{(1.5.12)}
\end{equation}
it follows that 
\begin{equation}
\lim_{\Lambda \nearrow \mathbb{Z}^{\nu }}U_{2(r-1),\Lambda }(0) \equiv 0
\quad \text{ for all } \quad\ r=3,4,5,\cdots  \label{(1.5.13)}
\end{equation}
and, consequently, 
\begin{equation}
\lim_{\Lambda \nearrow \mathbb{Z}^{\nu }}U_{2r,\Lambda }(0)\equiv 0 \quad 
\text{ for all }\quad\ r\geq 4  \label{(1.5.14)}
\end{equation}
\end{proposition}

\noindent \textit{Proof.} Write $2(r-1) = 2r-p+p-2$, with $2(2r-p)= 2r$,
which implies $r=p$. Then, by \eqref{(1.5.9)} and the Schwarz inequality, 
\begin{eqnarray*}
\left| \int x^{2(r-1)} dK_{\Lambda}(x) \right| & \le & \left( \int
x^{2(2r-p)} dK_{\Lambda}(x) \right)^{1/2} \left(\int x^{2(2p-2)}
dK_{\Lambda}(x) \right)^{1/2} \\
& = & \left( \int x^{2r} dK_{\Lambda}(x)\right)^{1/2} \left( \int x^{4(r-1)}
dK_{\Lambda}(x)\right)^{1/2}
\end{eqnarray*}
By assumption \eqref{(1.5.11)}, $\left|\displaystyle \int x^{4(r-1)}
dK_{\Lambda}(x)\right|$ is uniformly bounded in $\Lambda$, and the assertion
follows.

\hfill $\Box $

\begin{remark}
\label{Remark 1.5.1}

In \cite{MCJL2}, the authors suggest in their Theorem 6.2 that infinite
divisibility implies a Gaussian ch.f., under the sole assumption of the
existence of its thermodynamic limit. Proposition ~\ref{prop:1.5.1} proves,
however, the contrary, viz. that the ch.f. is Non-Gaussian for \emph{any}
finite volume, and, thus, Gaussian behavior may arise \emph{only} in the
thermodynamic limit. We show in the forthcoming Theorem \ref{th:2.3.1} that
this indeed happens, under certain assumptions.

The alternative proof sketched by the authors in a footnote, based on a
result of G.G. Hegerfeldt (\cite{Heg}; the reference given in \cite{MCJL2}
is incorrect), would be indeed valid if $\tilde{\Phi}_{\Lambda ,T}(t)$ were 
\emph{itself} infinitely divisible, not only its reciprocal \eqref{(1.5.6)}.
This is an open problem.
\end{remark}

\section{The connected two-point function ($r=2$) in the critical case}

\label{2}

\setcounter{equation}{0} \setcounter{Thm}{0}

\subsection{The framework}

\label{2.1}

We now turn to the framework introduced by one of us (M.R.) in \cite{MR},
which is complementary to that introduced by Verbeure and collaborators, of
which a comprehensive version is found in his book (\cite{Ver}, see, in
particular, Chap. 6, and the references given there). One different aspect
in \cite{MR}, which will be of central importance, is its use of a smoother
version of fluctuation operators and Fourier transform methods. For a
function $f$ defined on $\mathbb{Z}^{\nu }$, define its Fourier transform $
\hat{f}$ as (see \cite{BRo2}, p.253): 
\begin{equation}
f(x)=(2\pi )^{-\nu }\int_{\substack{ -\pi <k_{i}\leq \pi ;  \\ i=1,\ldots
,\nu }}d^{\nu }k\exp \left( ik\cdot x\right) \hat{f}(k)  \label{(2.1.4a)}
\end{equation}%
with 
\begin{equation}
\hat{f}(k)=\sum_{y\in \mathbb{Z}^{\nu }}\exp \left( -ik\cdot y\right) f(y)
\label{(2.1.4b)}
\end{equation}%
That is, the momenta $k$ are defined modulo the Brillouin zone $-\pi
<k_{i}\leq \pi ,i=1,\ldots ,\nu $. Let 
\begin{equation}
f_{R}(x)\equiv f(\left\vert x\right\vert /R)  \label{(2.1.5a)}
\end{equation}%
with 
\begin{equation}
f(s)=\left\{ 
\begin{array}{lll}
1 & \text{if} & 0\leq s\leq 1 \\ 
0 & \text{if} & s\geq 2%
\end{array}%
\right.  \label{(2.1.5b)}
\end{equation}%
be, for each $0<R<\infty $, a smooth (infinitely differentiable) function.
We shall choose for $\Lambda $ a ball centered at the origin with radius $R$%
, let $R\rightarrow \infty $, and use \eqref{(2.1.4a)}, \eqref{(2.1.4b)},
with $f\rightarrow f_{R}$, given by \eqref{(2.1.5a)} and \eqref{(2.1.5b)} as
the corresponding smooth approximation. It has the property (Lemma 2.6 of 
\cite{MR}): 
\begin{equation}
\hat{f_{R}}(k)=cR^{\nu }\hat{f}(Rk)  \label{(2.1.6)}
\end{equation}%
where $c$ is a positive constant. We omit the dependence of $c$ on $R$,
because it is uniformly bounded above and below by constants independent of $%
R$, the latter strictly positive (i.e. the r.h.s. of \eqref{(2.1.6)} is
asymptotically correct as $R\rightarrow \infty $): according to the
estimate, one of these two constants is employed. The smooth version of the
normalized fluctuation operator \eqref{(1.21)} will be denoted by $%
F_{R}^{\alpha }(A)$, where 
\begin{equation}
F_{R}^{\alpha }(A)=R^{-\alpha }\sum_{x\in \mathbb{Z}^{\nu }}\left(
A_{x}-\omega \left( A_{x}\right) \right) f_{R}(x)  \label{(2.1.7a)}
\end{equation}%
where the multiplicative constant $R^{-\alpha }$ corresponds to the
asymptotic (for $R\rightarrow \infty $) order of the variance \eqref{(1.22)}%
, with 
\begin{equation}
\alpha =\frac{\nu }{2}  \label{(2.1.7b)}
\end{equation}
in the normal case, and 
\begin{equation}
\alpha >\frac{\nu }{2}  \label{(2.1.7c)}
\end{equation}
in the anomalous case \eqref{(1.20)}. In \eqref{(2.1.7a)}, $A$ stands for
the observable, which will be $A=\sigma ^{3}$ in the Ising case, and the
reference to the thermal state $\omega _{T}$ is omitted, as in \eqref{(1.21)},
not to burden the notation.

We now follow \cite{MR}, Subsection 3.1. By translation invariance, the
connected correlation functions $W_{\omega _{T},C}^{r}(x_{1},\ldots ,x_{r})$
in \eqref{(2.1.1a)}, omitting now the reference to the temperature $T$ in $%
\omega $: 
\begin{equation}
W_{\omega ,C}^{r}(x_{1},\cdots ,x_{r})=W_{\omega ,C}^{r}(x_{1}-x_{2},\cdots
,x_{r-1}-x_{r})  \label{(2.1.8)}
\end{equation}
Introducing the variables 
\begin{equation}
y_{i}=x_{i}-x_{i+1}\ ,\qquad q_{i}=\sum_{j=1}^{i}k_{i}\ ,\quad i\leq r-1
\label{(2.1.9)}
\end{equation}
we now regard $W_{\omega ,C}^{r}$ as function of the variables $\{y_{i}\}$,
with $i\leq r-1$, and the Fourier transform $\widehat{W}_{\omega ,C}^{r}$ as
a function of the variables $q_{i};1\leq i\leq r-1$. We now pose

\emph{Assumption B} $W_{\omega ,C}^{r}$ is an $l^{1}$- function (i.e.,
absolutely summable) of the variables $y_{i}$, $i=1,\ldots ,r$. As a
consequence, $\widehat{W}_{\omega ,C}^{r}$ is a continuous and uniformly
bounded function of the variables $q_{i}$, $1\leq i\leq r-1$.

\subsection{The case $0<T<T_{c}$}

\label{2.2}

Theorem 3.3 of \cite{MR} may now be stated; we briefly sketch the proof for
the reader's convenience, because the argumements will be of major
importance when dealing with the critical point:

\begin{theorem}
\label{th:2.1.1}

Under Assumption B, 
\begin{equation}
0\leq \lim_{R\rightarrow \infty }\omega (F_{R}^{\nu /2}(A)^{r})<\infty \text{
\quad\ if \quad }r=2\ ,  \label{(2.1.10.1)}
\end{equation}%
whereas 
\begin{equation}
\lim_{R\rightarrow \infty }\omega (F_{R}^{\nu /2}(A)^{r})=0\text{ \quad\ if
\quad }r>2\ .  \label{(2.1.10.2)}
\end{equation}
\end{theorem}

\noindent \textit{Proof.} By \eqref{(2.1.6)} and the definition %
\eqref{(2.1.7a)}, 
\begin{equation}
\omega (F_{R}^{\nu /2}(A)^{r})=\mbox{ const. }R^{r\nu /2}I  \label{(2.1.11a)}
\end{equation}%
where 
\begin{equation}
I=\int_{\substack{ |q_{i}|\leq \pi ,  \\ i=1,\ldots ,r-1}}%
\prod_{i=1}^{r-1}dq_{i}\widehat{W}_{\omega ,C}^{r}(q_{1},\dots
,q_{r-1})J(\{Rq_{i}\})  \label{(2.1.11b)}
\end{equation}%
and 
\begin{equation}
J(\{Rq_{i}\})=\hat{f}(Rq_{1})\dots \hat{f}\left( -R(q_{1}+\cdots
+q_{r-1})\right)  \label{(2.1.11c)}
\end{equation}%
By \eqref{(2.1.11a)}, \eqref{(2.1.11b)} and \eqref{(2.1.11c)}, introducing
the change of variable $q^{\prime }=qR$, 
\begin{equation}
\omega (F_{R}^{\nu /2}(A)^{r})=\mbox{ const. }R^{r\nu /2}R^{-(r-1)\nu }\int 
_{\substack{ |q_{i}^{\prime }|\leq R\pi ;  \\ i=1,\ldots ,r-1}}%
J(\{q_{i}^{\prime }\})U(\{q_{i}^{\prime }\})  \label{(2.1.11d)}
\end{equation}%
where 
\begin{equation}
U(\{q_{i}^{^{\prime }}\})\equiv \widehat{W}_{\omega ,C}^{r}(q_{1}^{\prime
}/R,\ldots ,q_{r-1}^{\prime }/R)  \label{(2.1.11e)}
\end{equation}%
By assumption (see \eqref{(2.1.5a)}, \eqref{(2.1.5b)}), the $f$ functions
are smooth and of compact support. Therefore $\hat{f}(k)$ is uniformly
bounded in $k\in \mathbb{R}^{\nu }$ by a fixed function of fast decrease at
infinity in $k$-space. As a consequence, Assumption B, \eqref{(2.1.11a)} et
seq. and the Lebesgue dominated convergence theorem yield \eqref{(2.1.10.1)}
and \eqref{(2.1.10.2)}.

\hfill $\Box $

What about the applicability of Theorem \ref{th:2.1.1} to the Ising model?
We have:

\begin{theorem}
\label{th:2.1.2} If $A=\sigma ^{3}$ in \eqref{(2.1.7a)}, Assumption B holds
for the Ising model with n.n. interactions \eqref{(1.3.2)}. Consequently,
a.) \eqref{(2.1.10.1)} holds if $\omega =\omega _{T}$ is any ergodic thermal
state and $0\leq T<T_{c}$; b.) if $0\leq T\leq T_{0}<T_{c}$ and $T_{0}$ is
sufficiently small, \eqref{(2.1.10.2)} holds as well. In \eqref{(2.1.10.1)}, 
\begin{equation}
0<\lim_{R\rightarrow \infty }\omega _{T}(F_{R}^{\nu /2}(\sigma ^{3})^{2})
\label{(2.1.12a)}
\end{equation}%
if and only if 
\begin{equation}
0<T<T_{c}  \label{(2.1.12b)}
\end{equation}
\end{theorem}

\noindent \textit{Proof.} The first assertion follows from \cite{DCGR}, the
second from (\cite{BLOC}, Theorem 2, p. 431). By translation invariance of $%
\omega _{T}$ and Assumption B, 
\begin{equation*}
\lim_{R\rightarrow \infty }\omega _{T}(F_{R}^{\nu /2}(\sigma
^{3})^{2})=\sum_{x\in \mathbb{Z}^{\nu }}(\omega _{T}(\sigma _{x}^{3}\sigma
_{0}^{3})-\omega _{T}(\sigma _{0}^{3})^{2})
\end{equation*}%
By Griffiths' second inequality \eqref{(2.1.2)}, each summand above is
non-negative, and, therefore, restricting the sum to its value at $x=0$ we
obtain 
\begin{equation*}
\lim_{R\rightarrow \infty }\omega _{T}((F_{R}^{\nu /2}(\sigma
^{3})^{2}))\geq 1/4-\omega _{T}(\sigma _{0}^{3})^{2}
\end{equation*}%
Again by Griffiths' second inequality, the magnetization $\omega _{T}(\sigma
_{0}^{3})$ is a decreasing function of the temperature, which tends to $1/2$
as $T\rightarrow 0$, leading to \eqref{(2.1.12a)} and \eqref{(2.1.12b)}. The
\textquotedblleft only if\textquotedblright\ part follows from the fact that
for $T=0$ each ergodic component is an (infinite) product state and
therefore each term $\omega _{T}(\sigma _{x}^{3}\sigma _{0}^{3})-\omega
_{T}(\sigma _{0}^{3})^{2}$ in the first sum above is identically zero.

\hfill $\Box $

\begin{remark}
\label{Remark 2.1}

Case b.) was considerably extended by Newman (\cite{Ne1}, \cite{Ne2}) to the
whole region $0 \le T < T_{c}$, by a completely different and very elegant
approach.
\end{remark}

\subsection{The case of anomalous fluctuations: the critical (ergodic) state
for $r=2$ and $T=T_{c}$}

\label{2.3}

We now come back to \eqref{(2.1.7a)} and \eqref{(2.1.8)}, assuming
henceforth that $\omega =\omega _{T_{c}}$: 
\begin{equation}
\omega (F_{R}^{\alpha }(\sigma ^{3})^{2})=R^{-2\alpha }\sum_{x,y\in \mathbb{Z
}^{\nu }}f_{R}(\left\vert x\right\vert )f_{R}(\left\vert y\right\vert
)W_{\omega ,C}^{2}(\left\vert x-y\right\vert )  \label{(2.2.1)}
\end{equation}%
where, by translation invariance of $\omega $, 
\begin{equation}
W_{\omega ,C}^{2}(\left\vert x\right\vert )=\omega ((\sigma _{x}^{3}-\omega
(\sigma _{x}^{3}))(\sigma _{0}^{3}-\omega (\sigma _{0}^{3})))
\label{(2.2.2)}
\end{equation}%
The function $W_{\omega ,C}^{2}$ is a uniformly bounded function on $\mathbb{%
Z}^{\nu }$. Let $\mathbb{T}_{\nu }=(-\pi ,\pi ]^{\nu }$ be the $\nu $-
dimensional torus and $\mathcal{D}(\mathbb{T}_{\nu })$ denote the space of
all functions $\tilde{\phi}$ on $\mathbb{T}_{\nu }$ such that 
\begin{equation}
\tilde{\phi}\in C^{\infty }(\mathbb{R}^{\nu })\qquad \text{ with }\mathit{%
\qquad }\tilde{\phi}(k_{1},\ldots ,k_{n})=\phi (\exp (ik_{1}),\ldots ,\exp
(ik_{n}))  \label{(2.2.4)}
\end{equation}%
(see \cite{Ru}, p.190, exercise 22). If $\tilde{\phi}\in \mathcal{D}(\mathbb{%
T}_{\nu })$, then $\sum_{x\in \mathbb{Z}^{\nu }}(1+\left\vert x\right\vert
^{N})\left\vert \phi (x)\right\vert ^{2}<\infty $ for all integers $N<\infty 
$. These norms define a Fr\'{e}chet space topology on $\mathcal{D}(\mathbb{T}%
_{\nu })$ and any function such as $W_{\omega ,C}^{2}$ satisfying 
\begin{equation}
\left\vert W_{\omega ,C}^{2}(\left\vert x\right\vert )\right\vert \leq \text{
const. }(1+\left\vert x\right\vert )^{M}  \label{(2.2.3)}
\end{equation}%
(there is an obvious misprint in \cite{Ru}), for some positive integer $M$,
then it is defined as a distribution on $\mathbb{Z}^{\nu }$ by the equation: 
\begin{equation}
W_{\omega ,C}^{2}(\phi )\equiv \sum_{x\in \mathbb{Z}^{\nu }}W_{\omega
,C}^{2}(x)\phi (x)  \label{(2.2.5)}
\end{equation}%
This distribution is of \emph{positive type}, i.e., 
\begin{equation}
W_{\omega ,C}^{2}(\bar{\phi}_{-}\ast \phi )\geq 0\ ,\qquad \forall \phi \in 
\mathbb{Z}^{\nu }  \label{(2.2.6)}
\end{equation}%
where 
\begin{equation}
(f\ast g)(x)\equiv \sum_{y\in \mathbb{Z}^{\nu }}f(x-y)g(y)  \label{(2.2.7a)}
\end{equation}%
and 
\begin{equation}
\phi _{-}(x)\equiv \phi (-x)  \label{(2.2.7b)}
\end{equation}%
Notice that $W_{\omega ,C}^{2}$ satisfies \eqref{(2.2.3)}, being uniformly
bounded. By a simplified version of the Bochner-Schwartz theorem (Theorem
IX.10 of \cite{RSII}), and a simplified version of the argument at the
bottom of p.14 of \cite{RSII}, we find

\begin{theorem}
\label{th:2.2.1}$W_{\omega ,C}^{2}$ is the Fourier transform of a positive,
finite measure $\mu $ on the torus $\mathbb{T}_{\nu }$.
\end{theorem}

According to the above theorem, we may thus write 
\begin{equation}
W_{\omega ,C}^{2}(x)=(2\pi )^{-\nu }\int_{\mathbb{T}_{\nu }}d\mu (p)\exp
(ip\cdot x)  \label{(2.2.8)}
\end{equation}%
where $\mu $ is a positive, finite measure on $\mathbb{T}_{\nu }$, which has
the decomposition 
\begin{equation}
\mu =\mu _{\mathrm{a.c.}}+\mu _{\mathrm{p.p.}}+\mu _{\mathrm{s.c.}}
\label{(2.2.9)}
\end{equation}%
(\cite{RSI}, Theorems I.13 and I.14), where $\mu _{\mathrm{p.p.}}$ is pure
point, $\mu _{\mathrm{a.c.}}$ absolutely continuous and $\mu _{\mathrm{s.c.}%
} $ singular continuous with respect to Lebesgue measure on the torus. By
Theorem \ref{th:2.2.1}, $\mu _{\mathrm{a.c.}}$, $\mu _{\mathrm{p.p.}}$ and $%
\mu _{\mathrm{s.c.}}$ are all positive, finite measures on the torus.

In the proof of Theorem \ref{th:2.1.1}, we have used that $\widehat{W}
_{\omega ,C}^{2}$ is a continuous and uniformly bounded function on the
torus. Comparing with \eqref{(2.2.8)}, this means that 
\begin{equation}
d\mu (p) = \widehat{W}_{\omega ,C}^{2}(p)d^{\nu }p  \label{(2.2.10a)}
\end{equation}%
where 
\begin{equation}
\widehat{W}_{\omega ,C}^{2}(p)\text{ is a uniformly bounded continuous
function on }\mathbb{T}_{\nu }  \label{(2.2.10b)}
\end{equation}%
We now assume that this condition does \emph{not} hold, which implies that $%
W_{\omega ,C}^{2}$ is \emph{not} an $l^{1}$-function, in agreement with 
\emph{Criterion A}. We now have:

\begin{theorem}
\label{th:2.2.2}$\widehat{W}_{\omega ,C}^{2}$ is a positive definite
distribution on the torus $\mathbb{T}_{\nu }$. Write 
\begin{equation}
d\mu _{\mathrm{a.c.}}(p)=g(p)d^{\nu }p  \label{(2.2.11a)}
\end{equation}%
but, in contrast to \eqref{(2.2.10b)}, assume that 
\begin{equation}
g\text{ is not a uniformly bounded continuous function on }\mathbb{T}_{\nu }
\label{(2.2.11b)}
\end{equation}%
Then the observable $\sigma ^{3}$ exhibits anomalous fluctuations in the
state $\omega $, with, in \eqref{(1.20)}, 
\begin{equation}
\alpha =\nu /2+\rho /2  \label{(2.2.12a)}
\end{equation}%
where 
\begin{equation}
0<\rho <\nu  \label{(2.2.12b)}
\end{equation}%
If the above equations \eqref{(2.2.12a)} and \eqref{(2.2.12b)} hold, the
contributions of $\mu _{\mathrm{p.p.}}$ and $\mu _{\mathrm{s.c.}}$ vanish as 
$R\rightarrow \infty $.
\end{theorem}

\noindent \textit{Proof.} The first assertion follows from \eqref{(2.2.2)},
the definition \eqref{(2.2.6)} and the positivity of the state, i.e., 
\begin{equation}
\omega (A^{\ast }A)\geq 0\ ,\qquad \forall A\in \mathcal{A}  \label{(2.2.13)}
\end{equation}

We first assume that $\mu _{\mathrm{p.p.}}=0$ and $\mu _{\mathrm{s.c.}}=0$.

We may write the contribution of the absolutely continuous part %
\eqref{(2.2.10a)} to the r.h.s. of \eqref{(2.2.1)}, by the convolution
theorem as 
\begin{eqnarray}
I_{R} &\equiv &R^{-2\alpha }R^{\nu }\int_{\left\vert p\right\vert \leq \pi }%
\hat{f} (Rk)\hat{f}(-Rk)g(k)d^{\nu }k  \notag \\
&=&R^{-2\alpha }\int_{\left\vert k\right\vert \leq \pi R}\left\vert \hat{f}%
(k^{\prime })\right\vert ^{2}g(k^{\prime }/R)d^{\nu }k^{\prime }
\label{(2.2.14)}
\end{eqnarray}%
If $g$ is \emph{not} a uniformly bounded continuous function, it must
exhibit at some point of the torus (without loss, at the origin) an
integrable singularity, i.e., 
\begin{equation}
g(k^{\prime }/R)=O(\left( \left\vert k^{\prime }\right\vert /R\right)
^{-\rho })\qquad \text{as}\qquad \left\vert k^{\prime }\right\vert
\rightarrow 0  \label{(2.2.15)}
\end{equation}%
with \eqref{(2.2.12b)} valid, because the measure $\mu _{\mathrm{a.c.}}$ is
finite by Theorem \eqref{th:2.2.1}. By \eqref{(2.1.5a)} and \eqref{(2.1.5b)}%
, $\hat{f}$ is smooth and rapidly decreasing, and thus, by \eqref{(2.2.12b)}%
, the integral 
\begin{equation}
\int_{\mathbb{R}^{\nu }}\left\vert \hat{f}(k^{\prime })\right\vert
^{2}\left\vert k^{\prime }\right\vert ^{-\rho }d^{\nu }k^{\prime }<\infty
\label{(2.2.16)}
\end{equation}%
Let $\alpha $ be defined by \eqref{(2.2.12a)}. Then, by \eqref{(2.2.14)}, %
\eqref{(2.2.15)} and \eqref{(2.2.16)}, $\lim_{R\rightarrow \infty }I_{R}$
exists and, by the first inequality in \eqref{(2.2.12b)}, the observable $%
A=\sigma ^{3}$ has anomalous fluctuations in the state $\omega $, according
to the definition \eqref{(1.20)}.

In order to show the last assertion, observe that the contributions of $\mu
_{\mathrm{p.p.}}$ and $\mu _{\mathrm{s.c.}}$ to the r.h.s. of \eqref{(2.2.1)}
equal 
\begin{equation*}
\lim_{R\rightarrow \infty }R^{-2\alpha }R^{\nu }\int_{|k|\leq \pi
}\left\vert \hat{f}(Rk)\right\vert ^{2}\left( d\mu _{\mathrm{p.p.}}(k)+d\mu
_{\mathrm{s.c.}}(k)\right) =0
\end{equation*}%
by \eqref{(2.1.12a)}, \eqref{(2.1.12b)}, and the bound 
\begin{equation*}
\left\vert \hat{f}(Rk)\right\vert \leq \mbox{ const. }
\end{equation*}%
with the constant above independent of $R$, which results from %
\eqref{(2.1.6)}.

\hfill $\Box $

Let $f\in C_{0}^{\infty }(\mathbb{R}^{\nu })$. Then (\cite{LL}, Theorem 5.9:
Fourier transform of $\left\vert x\right\vert ^{\rho -\nu }$), with $c_{\rho
}\equiv \pi ^{-\rho /2}\Gamma (\rho /2)$, 
\begin{equation}
c_{\rho }(\left\vert k\right\vert ^{-\rho }\hat{f}(k))^{\vee }(x)=c_{\nu
-\rho }\int_{\mathbb{R}^{\nu }}|x-y|^{\rho -\nu }f(y)dy  \label{(2.2.19)}
\end{equation}%
where the superscript $^{\vee }$ denotes the inverse of the Fourier
transform. We have, however, replaced their $\alpha $ by our $\rho $, which
satisfies \eqref{(2.2.12b)}, i.e., their assumption. Our conventions for the
Fourier transform differ from those in \cite{LL}, so that the above formulas 
may differ from those in our notation by constants which do not alter the final
statement. If we now set 
\begin{equation}
\rho =2-\eta  \label{(2.2.20)}
\end{equation}%
we obtain that 
\begin{equation}
W_{\omega ,C}^{2}(|x|)=\left\vert x\right\vert ^{-(\nu -2+\eta )}
\label{(2.2.21)}
\end{equation}%
satisfies \eqref{(2.2.19)}. For $\rho >\nu /2$, the r.h.s. of %
\eqref{(2.2.19)} is not in $L^{p}(\mathbb{R}^{\nu })$ for any $p\leq 2$, as
the authors remark p. 131 of \cite{LL}, which makes inclusion of an $f\in
C_{0}^{\infty }(\mathbb{R}^{\nu })$ in \eqref{(2.2.19)} necessary. In the
lattice case these considerations become simplified, but \eqref{(2.2.15)}
shows that only the asymptotic behavior at large distances is an issue, and,
accordingly, we replace \eqref{(2.2.21)} by the

\emph{Standard Assumption} (\cite{GML}, \cite{MCJL1}) 
\begin{equation}
W_{\omega ,C}^{2}(|x|)=O(\left\vert x\right\vert ^{-(\nu -2+\eta )})\qquad 
\text{ as \qquad }\left\vert x\right\vert \rightarrow \infty
\label{(2.2.22)}
\end{equation}

\begin{remark}
\label{Remark 2.2}

The standard assumption is proved true if $\nu=2$ by a closed-form
calculation \cite{MW}, yielding 
\begin{equation}  \label{(2.2.23)}
\eta = 0.25
\end{equation}
\end{remark}

The following theorem is therefore complementary to Theorem \ref{th:2.2.2}:

\begin{theorem}
\label{th:2.2.3} Under the Standard Assumption \eqref{(2.2.22)}, $W_{\omega
,C}^{2}(|x|)$ is not an $l^{1}$- function, and the observable $A=\sigma ^{3}$
has anomalous fluctuations in the state $\omega $, according to definition %
\eqref{(1.20)}.
\end{theorem}

\noindent \textit{Proof.} By \eqref{(2.2.22)} and Theorem 17.8 by Glimm and
Jaffe \cite{GJ}, 
\begin{equation}
\eta \leq 1  \label{(2.2.24)}
\end{equation}%
and hence, by \eqref{(2.2.22)}, $W_{\omega ,C}^{2}(|x|)$ is not an $l^{1}$-
function. By \eqref{(2.2.22)} and Griffiths' second inequality %
\eqref{(2.1.2)}, there exists a $R_{0}<\infty $ and two constants $c>0$ and $%
d<\infty $ such that 
\begin{equation}
c~\left\vert x\right\vert ^{-(\nu -2+\eta )}\leq W_{\omega ,C}^{2}(|x|)\leq
d~\left\vert x\right\vert ^{-(\nu -2+\eta )}\qquad \mbox{ for all }\qquad
\left\vert x\right\vert \geq R_{0}  \label{(2.2.25)}
\end{equation}%
We now write 
\begin{equation}
\alpha = (\nu +\rho ) /2  \label{(2.2.26)}
\end{equation}%
By \eqref{(2.2.1)} and \eqref{(2.2.26)}, 
\begin{equation}
\omega (F_{R}^{\alpha }(\sigma ^{3})^{2})=R^{-\nu }R^{-\rho }\sum_{x,y\in 
\mathbb{Z}^{\nu }}f_{R}(\left\vert x\right\vert )f_{R}(\left\vert
y\right\vert )W_{\omega ,C}^{2}(\left\vert x-y\right\vert )  \label{(2.2.27)}
\end{equation}%
By \eqref{(2.1.5a)}, \eqref{(2.1.5b)}, \eqref{(2.2.24)}, \eqref{(2.2.25)}, %
\eqref{(2.2.27)} and Griffiths' second inequality \eqref{(2.1.2)}, we obtain 
\begin{eqnarray}
\omega ((F_{R}^{\alpha }(\sigma ^{3})^{2})) &\geq &cR^{-\nu
}\int_{R_{0}}^{2R}drr^{\nu -1}r^{-(\nu -2+\eta )}+o\left( R\right)  \notag \\
&=&cR^{-\rho }[(2R)^{2-\eta }-R_{0}^{2-\eta }]+o\left( R\right)
\label{(2.2.28)}
\end{eqnarray}%
By \eqref{(2.2.24)}, choosing 
\begin{equation}
\rho =2-\eta  \label{(2.2.29)}
\end{equation}%
the $R_{0}$- term in \eqref{(2.2.28)} does not contribute in the limit $%
R\rightarrow \infty $, and we obtain from \eqref{(2.2.28)} 
\begin{equation}
\lim_{R\rightarrow \infty }\omega ((F_{R}^{\alpha }(\sigma ^{3})^{2}))\geq
c~2^{2-\eta }  \label{(2.2.30a)}
\end{equation}%
The second inequality in \eqref{(2.2.25)} yields similarly, with $\alpha $
and $\rho $ as in \eqref{(2.2.26)} and \eqref{(2.2.29)}, 
\begin{equation}
\lim_{R\rightarrow \infty }\omega ((F_{R}^{\alpha }(\sigma ^{3})^{2}))\leq
d~2^{2-\eta }  \label{(2.2.30b)}
\end{equation}%
By \eqref{(2.2.30a)}, \eqref{(2.2.30b)} ,\eqref{(2.2.26)} and %
\eqref{(2.2.29)}, the second assertion of the theorem follows.

\hfill $\Box $

\begin{remark}
\label{Remark 2.3}

Equation \eqref{(2.2.24)} was proved in \cite{GJ}, Theorem 17.8, under the 
\emph{sole} assumption \eqref{(2.2.22)}. It uses in an essential way the
Lebowitz inequality \eqref{(2.1.3)}, as well as the Simon-Lieb correlation
inequality (\cite{Simon}, \cite{Lieb}).
\end{remark}

\begin{remark}
\label{Remark 2.4}

Theorems \ref{th:2.2.2} and \ref{th:2.2.3} together make Criterion A very
plausible. Assumption \eqref{(2.2.11b)} implies that $W_{\omega ,C}^{2}(|x|)$
is not an $l^{1}$- function, and has, at the same time, an integrable
singularity in its Fourier transform, which is related, in a certain precise
sense previously explained, to the asymptotic form \eqref{(2.2.22)}. On the
other hand, if \eqref{(2.2.22)} is assumed, then $W_{\omega ,C}^{2}(|x|)$ is
not an $l^{1}$- function, and the observable $A=\sigma ^{3}$ has anomalous
fluctuations in the state $\omega $, according to Definition \eqref{(1.20)}.

The above remark refers to the first inequality in \eqref{(2.2.12b)} (given
\eqref{(2.2.12a)}). It is also important to notice that the second
inequality in \eqref{(2.2.12b)} also has a physical-mathematical content
which is natural in the present approach. Indeed, setting \eqref{(2.2.20)}
into \eqref{(2.2.12b)}, we find the requirement 
\begin{equation*}
2-\eta < \nu
\end{equation*}
or 
\begin{equation}  \label{(2.2.31)}
\nu-(2-\eta) > 0
\end{equation}
Comparing with \eqref{(2.2.22)}, we see that \eqref{(2.2.31)} is just the
requirement of ergodicity \eqref{(1.2.6)} of the state $\omega$, which is a
general result for quantum spin systems by \cite{BRo1}, Example 4.3.24,
p.401.
\end{remark}

\subsection{The connected $r$-point functions in the critical case ($r\geq 4$
) and the quasi-free nature of $\protect\omega _{T}^{F}$}

\label{2.4}

We now come back to \eqref{(2.1.11d)} and \eqref{(2.1.11e)}, and pose, for
all $r$, $A=\sigma ^{3}$, 
\begin{equation}
\omega _{T_{c}}(F_{R}^{\nu /2+\rho /2}(\sigma ^{3})^{r})=\text{const. }
R^{r\nu /2}R^{-(r-1)\nu }R^{\rho ^{\prime }(r-2)}  \label{(2.3.1)}
\end{equation}%
where the \textquotedblleft constant\textquotedblright\ is to be understood
as in \eqref{(2.1.6)} et seq.. In \eqref{(2.3.1)} we introduce a new
exponent $\rho ^{\prime }$, for $r\geq 4$. This Ansatz has the general form
of the scaling hypothesis VI, p.434, in \cite{GML}:

\emph{Scaling Assumption C} 
\begin{equation}
\widehat{W}_{\omega _{T_{c}},C}^{r}(q_{1}^{\prime }/R,\ldots
,q_{r-1}^{\prime }/R)= R^{\rho ^{\prime }(r-2)}\widehat{W}_{\omega
_{T_{c}},C}^{r}(q_{1}^{\prime },\ldots 
,q_{r-1}^{\prime }) \qquad \text{ for \qquad }
r\geq 4  \label{(2.3.2)}
\end{equation}
with an exponent $\rho ^{\prime }$ satisfying 
\begin{equation}
0<\rho ^{\prime }<\infty  \label{(2.3.3)}
\end{equation}

$\widehat{W}_{\omega _{T_{c}},C}^{r}$ is, in fact, a (Schwartz)
distribution defined by
\begin{equation*}
\langle \widehat{W}_{\omega _{T_{c}},C}^{r}, \hat{f} \rangle = \langle
W_{\omega   _{T_{c}},C}^{r},f \rangle \equiv 
\sum_{x_{1}, \ldots x_{r} \in \mathbb{Z}^{\nu}} W_{\omega
  _{T_{c}},C}^{r}(x_{1} \ldots x_{r})f(x_{1}, \ldots, x_{r}) 
\end{equation*}
for functions $f$ of bounded support. Equation \eqref{(2.3.2)}
corresponds to defining the scaled distribution 
$\widehat{W}_{\omega _{T_{c}},C}^{r,R}$ associated to
$\widehat{W}_{\omega _{T_{c}},C}^{r}$ by 
$$
\langle \widehat{W}_{\omega _{T_{c}},C}^{r,R}, \hat{f} \rangle
\equiv \langle \widehat{W}_{\omega _{T_{c}},C}^{r},\hat{f}_{R} \rangle
$$ 
where
$$
\hat{f}_{R}(q_{1}^{\prime },\ldots q_{r-1}^{\prime }) \equiv
R^{\nu(r-1)} f(R q_{1}^{\prime },\ldots,R q_{r-1}^{\prime }) 
$$
with $|q_{i}^{\prime }| \le \pi, i=1, \cdots, r-1$. Finally, equation
\eqref{(2.3.2)} should be written precisely as 
$$
\langle \widehat{W}_{\omega _{T_{c}},C}^{r},\hat{f}_{R}\rangle =
R^{\rho^{\prime}(r-2)} \langle \widehat{W}_{\omega _{T_{c}},C}^{r},
\hat{f} \rangle 
$$

Note also that 
the condition $\rho ^{\prime }>0$ corresponds, for $r\geq 4$, to
the analogue of the first of \eqref{(2.2.12b)} for $r=2$. For $r=2$ in 
\eqref{(2.3.1)}, the r.h.s. of \eqref{(2.3.1)} is $O(1)$ by Theorems 2.4 and
2.5.

\begin{theorem}
\label{th:2.3.1}

Assume that 
\begin{equation}
\text{ there exists }\lim_{R\rightarrow \infty }\omega _{T_{c}}(F_{R}^{\nu
/2+\rho /2}(\sigma ^{3})^{r})\quad \text{ for \quad }r\geq 4  \label{(2.3.4)}
\end{equation}
and, moreover, the limit agrees with its LLRO version, that is to say, one
may replace in \eqref{(2.3.4)}, $\omega _{T_{c}}$ by $\omega _{R,T_{c}}$.
Then, under the Scaling Assumption C, 
\begin{equation}
\lim_{R\rightarrow \infty }\omega _{T_{c}}((F_{R}^{\nu /2+\rho /2}(\sigma
^{3})^{r})=0\quad \text{ for }\quad r\geq 4  \label{(2.3.5)}
\end{equation}
and thus the fluctuation state $\omega _{T_{c}}^{F}$ is quasi-free.
\end{theorem}

\noindent \textit{Proof.} By \eqref{(2.3.1)} and Scaling Assumption C, 
\begin{equation}
\omega _{T_{c}}(F_{R}^{\nu /2+\rho /2}(\sigma ^{3})^{r})=\text{const. }
R^{r(\rho ^{\prime }-\nu /2-\rho /2)}R^{\nu -2\rho ^{\prime }}
\label{(2.3.6)}
\end{equation}
Assume, first, that 
\begin{equation}
0<\rho ^{\prime }<\alpha = (\nu + \rho )/2  \label{(2.3.7a)}
\end{equation}
Then, by \eqref{(2.3.6)} and \eqref{(2.3.7a)}, for $r$ sufficiently large, 
\begin{equation}
\lim_{R\rightarrow \infty }\omega _{T_{c}}(F_{R}^{\nu /2+\rho /2}(\sigma
^{3})^{r})=0  \label{(2.3.7b)}
\end{equation}
and, hence, by the assumption of agreement with the LLRO version, 
\begin{equation}
\lim_{R\rightarrow \infty }\omega _{R,T_{c}}(F_{R}^{\nu /2+\rho /2}(\sigma
^{3})^{r})=0\mbox{ for sufficiently large }r  \label{(2.3.7c)}
\end{equation}
It follows from \eqref{(2.3.7c)} and Proposition ~\eqref{prop:1.5.2} that 
\begin{equation}
\lim_{R\rightarrow \infty }\omega _{R,T_{c}}(F_{R}^{\nu /2+\rho /2}(\sigma
^{3})^{r})=0\mbox{ for all }r\geq 4  \label{(2.3.8)}
\end{equation}%
and therefore the fluctuation state is quasi-free. If, on the other hand, 
\begin{equation}
\rho ^{\prime }=\alpha  \label{(2.3.9)}
\end{equation}
then $\nu -2\rho ^{\prime }=2(\nu /2-\rho ^{\prime })=-\rho <0$ and 
\eqref{(2.3.5)} holds as well. The last case, $\rho >\alpha $ implies that
the r.h.s of \eqref{(2.3.6)} diverges, as $R\rightarrow \infty $, for
sufficiently large $r$, contradicting the assumption \eqref{(2.3.4)}.

\hfill $\Box $

\begin{remark}
\label{Remark 2.3.1} It is the necessity of using a definite normalization
for all $r$ in \eqref{(2.3.4)} -- the only one which guarantees the
existence of the thermodynamic limit for $r=2$, due to the existence of
anomalous fluctuations -- which leads to a quasi-free fluctuation state in
the above theorem; in particular, the fact that the case of equality 
\eqref{(2.3.9)} satifies \eqref{(2.3.5)}.
\end{remark}

\begin{remark}
\label{Remark 2.3.2}

By Theorem \ref{th:2.3.1} and Proposition \ref{prop:1.5.2} the fluctuation
state undergoes a basic structural change, from interacting to quasi-free,
in the passage to the thermodynamic limit. This is different from previous
references to structural changes of states, which concerned states of \emph{
infinite} systems: one exception is the change from pure to mixed state in
the quantum theory of measurement in \cite{Wre3}, which also occurred in the
passage to the thermodynamic limit. Both transformations, therefore, occur
naturally, and should be accounted for in a general theory, see
Section \ref{4}. 
\end{remark}

\section{Long-range interactions: universality and its violations\label{3}}

\setcounter{equation}{0} \setcounter{Thm}{0}

It is well-known that the critical exponents or indices depend only on
rather general qualitative features of physical systems, such as
dimensionality, symmetry and range of interaction: this is the universality
of critical phenomena \cite{Fis}. We shall be concerned only with the
Ising universality class, which concerns only a discrete symmetry (the spin-reflection
symmetry), but consider in this section long-range interactions.
Indeed, the short vs. long range character of the interaction may be of decisive
importance in determining whether a phase transition of a given type
(e.g., first-order) takes place in a system, e.g. for Ising systems in
dimension $\nu=1$ no first-order phase transition is possible if the
interaction is short-range; in order to have one, we must consider
long-range interactions, to which we now turn.

\subsection{Brief review of results and problems}

Concerning long-range interactions, we shall limit ourselves to a prototype,
the Dyson \cite{Dy} model - referred to as $D_{\alpha }$-model. It is given
by \eqref{(1.2.1)} with $\delta =0$ and 
\begin{equation}
j(\left\vert x\right\vert )=\frac{1}{\left\vert x\right\vert ^{\alpha }}
\label{(3.1.3)}
\end{equation}
in dimension 
\begin{equation}
\nu =1  \label{(3.1.4)}
\end{equation}
For 
\begin{equation}
1<\alpha <2  \label{(3.1.5)}
\end{equation}
it was proved by Dyson that the model exhibits a ferromagnetic phase
transition (for an independent proof, see \cite{FILS}). It has been proved
(see \cite{AF}, Prop. 2.1) that several critical exponents, including $\beta 
$ and $\gamma $ in \eqref{(1.16)} and \eqref{(1.17)}, assume their
mean-field values, viz. 
\begin{equation}
\beta =\frac{1}{2}  \label{(3.1.6a)}
\end{equation}
and 
\begin{equation}
\gamma =1  \label{(3.1.6b)}
\end{equation}
(there are others, see \cite{AF}), as soon as 
\begin{equation}
1<\alpha \leq \frac{3}{2}  \label{(3.1.7)}
\end{equation}
This is not quite \eqref{(3.1.5)}, but it is believed that the result is
correct,
in agreement with the
predictions of Bleher and Sinai (\cite{BS1}, \cite{BS2}, see also \cite{Sew1},
Sec. 5.4.3 for a pedagogic outline) for the Dyson hierarchical model. It is of interest
that Lohmann, Slade and Wallace \cite{LSW} have proved that the exponent $\eta$ keeps its
mean-field value even for $\alpha > 3/2$.
We come back in Section \ref{4} to a more detailed analysis of the
predictions of Bleher and Sinai. 

This universality -- the independence of the critical indices on the rate $
\alpha $ of fall-off of the potential \eqref{(3.1.3)} -- need not happen for
all \textquotedblleft critical indices\textquotedblright , in particular not
for those related to the \emph{dynamics}. This issue -- the breaking of
universality for \textquotedblleft dynamical critical
indices\textquotedblright\ -- has been seldom studied: it is the subject of
the next subsection.

\begin{remark}
\label{remark:3.1}
Regarding the interval \eqref{(3.1.7)} in which critical exponents
$\beta $ and $\gamma $ assume their mean-field values \eqref{(3.1.6a)}
and \eqref{(3.1.6b)}, it seems, at first sight, that no dimension may
be attached to the power 
decay $\alpha $ of the interaction in $D_{\alpha }$-model. However, since
the hierarchical model has been introduced by Dyson to establish long range
order in the $D_{\alpha }$-model; since the spectral dimension $d$ of the
hierarchical Laplacean $\Delta ^{\mathrm{hier.}}$ (see e.g. \cite{BAHM})
is able to distinguish whether a random walk generated by $\Delta ^{
\mathrm{hier.}}$ is recurrent if $d\leq 2$ or transient if $d>2$; the
function $\alpha (d)=(d+2)/d$ deduced from these facts establishes that $
\alpha (\infty )=1$, $\alpha (2)=2$ and $\alpha (4)=3/2$. So,
$1<\alpha \left(
d\right) <2$ as $d\in \left( 2,\infty \right) $ and the interval \eqref{(3.1.7)}
refers to $d\geq 4$, the range for which the fluctuation in the Ising model
at critical point are Gaussian and the critical exponents are classical.
See also \cite{ADCV} Proposition 1.3, Corollaries 1.4 and 1.5 and
references given there for a closely related viewpoint using infrared bound
and substantial argument based on the continuity of order parameter.
\end{remark}

\subsection{Aspects of dynamics: violations of universality in problems
  of energetic and dynamic stability}
\label{3.2}

Up to now we have made little reference to the subject of \emph{dynamics}.
The dynamics of fluctuations is of more importance as in the work of Wehrl 
\cite{Wehrl} which proved the important fact that the \textquotedblleft
true\textquotedblright\ fluctuations (i.e., corresponding to zero momentum)
do not satisfy the KMS condition, suggesting the quasi-free nature of the
fluctuation states for all temperatures (see Remark \ref{Remark 1.4.1}). How
do the other macroscopic observables -- the intensive observables -- behave
in this connection, i.e., in a non-equilibrium framework?

In this section we consider the dynamics of intensive observables, viz. the
mean magnetization. It is shown that this dynamics is able to distinguish
the two universality classes -- short range and long-range -- through the
different rates of decay of unstable states in the system, viz. macroscopic
\textquotedblleft liquid drops\textquotedblright . Even more interestingly,
these effects are able to distinguish the several types of (nonuniversal)
behavior within one class (e.g., a class of long-range interactions), a
topic which has been very seldom touched upon.

A heuristic, but illuminating way to understand the difference of behavior
of short and long-range interactions was proposed by Haag in his seminal
paper on the mathematical structure of the BCS model \cite{Haag}. Consider,
for simplicity, the ground state of a ferromagnet (e.g., Ising or
Heisenberg) with spins oriented along the $+3$-direction (say). He argued
that the energy necessary to build a \textquotedblleft liquid
droplet\textquotedblright\ of oppositely oriented spins (along the $-3$%
-direction) of large volume $V$ is of larger order in $V$ for long-range
interactions than for those of short range. Thus, long-range interactions 
\emph{stabilize} the system: we shall call this \textquotedblleft
freezing\textquotedblright\ property of long-range interactions
\textquotedblleft Haag's principle\textquotedblright . When applying Haag's
principle to particular models, it will be possible to make more precise the
conditions of its validity.

A prototype of models of long-range interactions is Dyson's $D_{\alpha }$
model. The energy $E$ necessary to create a negatively polarized droplet of
length $N$ in a positively polarized state equals (see also \cite{Sew1},
p.126, which we follow here): 
\begin{equation}
E=\text{const.}N^{2-\alpha }  \label{(3.2.1)}
\end{equation}%
In the critical region \eqref{(3.1.5)}, this energy is of an order of
magnitude greater than that which is necessary for short range interactions
(e.g., bounded by a constant in the case of finite range). This is an
expression of Haag's principle for the Dyson model, whose validity is,
however, restricted to the values of $\alpha $ corresponding to the
phase-transition region. This is not an accident: \eqref{(3.2.1)} is the
central ingredient in an energy-entropy estimate which may even be used to 
\emph{prove} the existence of a phase transition in the model (see
\cite{Sew1}, p. 126 and references given there). A proof of phase
transitions in the Dyson model along these lines was given in \cite{CFMP}.

If one considers now a state
consisting of a large droplet of spins polarized along the $1$-direction,
immersed in a background of spins polarized along the $3$-direction, the
energy fluctuation $\Delta E$ of the Dyson Hamiltonian in this state is of
the order of the r.h.s. of \eqref{(3.2.1)}. This state will be unstable and
decay with a half-life $\Gamma $ given by the time-energy uncertainty
principle $\Delta E\cdot \Gamma \approx \hbar $ (in the case of a Markovian
approach to equilibrium , i.e., of type $\exp (-t/\Gamma )$), otherwise,
instead of $\Gamma $, one would have a quantity called \textquotedblleft
sojourn time\textquotedblright\ in the state (see, e.g., \cite{MWB}, Theorem
3.17, p. 81, and references given there). From this heuristics and 
\eqref{(3.2.1)}, we are led to conclude: the rate of decay of such a state
is the greater, the longer the range of interaction, i.e, the closer the
parameter $\alpha $ is to one, inside the open interval $(1,2)$. We shall
call this heuristic principle, for brevity, \textquotedblleft Haag's
principle\textquotedblright , because it is a direct consequence of the rise
of energetic stability of large droplets when the range of interaction
increases. If, in \eqref{(1.2.1)}, we consider the Ising limit, i.e., set $%
\delta \equiv 0$, we obtain the generalized Ising model (gIm) studied by
Emch \cite{Em} and Radin \cite{Ra} (we changed the overall sign to agree
with Radin's notation) 
\begin{equation}
H_{\Lambda }=\frac{1}{2}\sum_{x,y\in \Lambda }j(|x-y|)\sigma _{x}^{3}\sigma
_{y}^{3}  \label{(3.2.2)}
\end{equation}

For a class of initial states, characterized in \cite{Ra}, of which the
product states 
\begin{equation}
\omega _{p,q}\equiv \prod_{x\in \mathbb{Z}^{\nu }}\left( p~\left\vert \pm
\right) _{x}+q~\left\vert \pm \right) _{x}^{\prime }\right)   \label{(3.2.3)}
\end{equation}%
are prototypes, where $p$ and $q$ are (without loss) non-negative and 
\begin{equation}
\sigma _{x}^{1}~\left\vert \pm \right) _{x}=\pm ~\left\vert \pm \right)
_{x}\qquad \text{ and\qquad\ }\sigma _{x}^{2}~\left\vert \pm \right)
_{x}^{\prime }=\pm ~\left\vert \pm \right) _{x}^{\prime }\qquad \text{ and
\qquad }p^{2}+q^{2}=1  \label{(3.2.4)}
\end{equation}%
the approach to equilibrium is explicit \cite{Em}: 
\begin{equation}
\omega _{p,q}(\tau _{t}(\sigma _{x}^{1}))=\pm p\prod_{y\in \mathbb{Z}^{\nu
}}\cos ^{2}(2J(\left\vert y\right\vert )t)\qquad \text{ for any }x\in 
\mathbb{Z}^{\nu }  \label{(3.2.5)}
\end{equation}%
with a corresponding relation for the $2$-component. By the stability
condition \eqref{(1.2.2)},  
\begin{equation*}
\lim_{\left\vert x\right\vert \rightarrow \infty }j(\left\vert x\right\vert
)=0\,,
\end{equation*}%
and, in addition, the above infinite product converges absolutely. Let, now, 
$\nu =1$, and define a subclass of models of the gIm:

For Dyson's $D_{\alpha }$ model, the infinite product in \eqref{(3.2.5)},
relevant to the decay of the magnetization along the $1$- or $2$-axis, is
given by 
\begin{equation}
\mathrm{Cl}_{1;\alpha }(t)\equiv \prod_{j\in \mathbb{N}}\cos (j^{-\alpha }t)
\label{(3.2.6)}
\end{equation}%
where $\mathrm{Cl}$ denotes the Cloitre function introduced in \cite{AlKi}.
We have the following lemma due to Radin (\cite{Ra}, Lemma4, p. 2954), whose
proof we reproduce for the reader's convenience, because it
\textquotedblleft explains\textquotedblright\ part of the forthcoming
theorem of Albert and Kiessling (Theorem 5.1 of \cite{AlKi}):

\begin{lemma}
\label{lem:3.2.2}If $\alpha >1$, $\exists ~c>0$ such that 
\begin{equation}
\left\vert \mathrm{Cl}_{1;\alpha }(t)\right\vert \leq \exp (-c|t|^{1/\alpha
})  \label{(3.2.7)}
\end{equation}
\end{lemma}

\noindent \textit{Proof.} For $0<x<1$, we have $0<\cos x<1-cx^{2}$, for some 
$c>0$, and $1-x\leq \exp (-x)$. Therefore, for $t>0$, we have 
\begin{eqnarray*}
|\mathrm{Cl}_{1;\alpha }(t)| &\leq &\prod\nolimits_{j>t^{1/\alpha
}}\left\vert \cos (t/j^{\alpha })\right\vert  \\
&\leq &\prod\nolimits_{j>t^{1/\alpha }}\left( 1-ct^{2}/j^{2\alpha }\right) 
\\
&\leq &\prod\nolimits_{j>t^{1/\alpha }}\exp \left( -ct^{2}/j^{2\alpha
}\right) =\exp \left( -c\sum\nolimits_{j>t^{1/\alpha }}t^{2}/j^{2\alpha
}\right) 
\end{eqnarray*}

By integral approximation, $\dsum_{j>t^{1/\alpha }}t^{2}/j^{2\alpha }\geq
t^{1/\alpha }$, and transition to negative $t$ finally yields the full
result.

\hfill $\Box $

The following special case of Theorem 5.1 of \cite{AlKi} considerably
improves Lemma \ref{lem:3.2.2}:

\begin{theorem}
\label{th:3.2.3} Let 
\begin{equation}
\alpha >\frac{1}{2}  \label{(3.2.8)}
\end{equation}%
Then, $\forall t\in \mathbb{R}$, 
\begin{equation}
\mathrm{Cl}_{1;\alpha }(t)=\exp (-C_{1;\alpha }|t|^{1/\alpha })F_{1;\alpha
}(|t|)  \label{(3.2.9)}
\end{equation}%
where 
\begin{equation}
\left\vert F_{1;\alpha }(|t|)\right\vert \leq \exp (K_{1;\alpha }\left\vert
t\right\vert ^{1/(\alpha +1)})\ ,\qquad \text{ for some constant }%
K_{1;\alpha }>0  \label{(3.2.10)}
\end{equation}%
and where 
\begin{equation}
C_{1;\alpha }=-\frac{1}{\alpha }\int_{0}^{\infty }\log \left\vert \cos \xi
\right\vert \frac{1}{\xi ^{1+1/\alpha }}d\xi   \label{(3.2.11)}
\end{equation}
\end{theorem}

The proof of the above theorem in \cite{AlKi} follows in part Theorem 1 of 
\cite{Ki}. Since increasing the range of interaction in the Dyson model in
the range \eqref{(3.1.5)} means varying $\alpha$ from $2$ to $1$, in the
open interval, we have the following

\begin{corollary}
\label{cor:3.2.1} 
\begin{equation}
\left\vert \omega _{p,q}(\tau _{t}(\sigma _{x}^{1}))\right\vert =O(\exp
(-c\left\vert t\right\vert ^{1/\alpha })\ ,\qquad \text{for some }c>0\text{
as }\left\vert t\right\vert \rightarrow \infty \ .  \label{(3.2.12)}
\end{equation}%
\eqref{(3.2.12)} defines a \textquotedblleft dynamical critical
exponent\textquotedblright\ $1/\alpha $, which satisfies Haag's principle
and violates universality.
\end{corollary}

The question may now be posed whether the mean-field limit describes the
decay of the states $\omega_{p,q}$ for large range of interaction.

We now show that there exists a (trivial) exact solution of the non-linear
system derived in Appendix A of \cite{LvH}, of which a special case is the
mean-field limit of the model of the previous section, with the initial
state \eqref{(3.2.3)}, \eqref{(3.2.4)} and that, as a consequence, the
mean-field limit does not describe the decay of the states $\omega_{p,q}$
for large range of interaction.

We consider the (mean-field) Hamiltonian 
\begin{equation}
H_{\Lambda }^{\mathrm{m.f.}}\equiv \frac{1}{|\Lambda |}\sum_{x,y\in \Lambda
}H^{\mathrm{m.f.}}(x,y)  \label{(3.2.13)}
\end{equation}%
where 
\begin{equation}
H^{\mathrm{m.f.}}(x,y)=a\sigma _{x}^{1}\sigma _{y}^{1}+b\sigma
_{x}^{2}\sigma _{y}^{2}+c\sigma _{x}^{3}\sigma _{y}^{3}  \label{(3.2.14)}
\end{equation}%
The above corresponds to the Hamiltonian (3.2.1) for a van der Waals spin
system in reference \cite{LvH}. We define, as in (A.1) of (\cite{LvH},
Appendix A), the time-dependent spin operator (changing to the notation $%
S(x;t)\rightarrow \sigma (x;t)$) 
\begin{equation}
\sigma ^{i}(x;t)\equiv \exp (iH_{\Lambda }t)\sigma ^{i}(x;0)\exp
(-iH_{\Lambda }t)\qquad \text{ for }i=1,2,3  \label{(3.2.15)}
\end{equation}%
and, as in \cite{LvH}, the time-dependent space-average of $\sigma
^{i}(x;t),i=1,2,3$ by 
\begin{equation}
\eta _{\Lambda }(\sigma ^{i};t)\equiv \frac{1}{|\Lambda |}\sum_{x\in \Lambda
}\sigma ^{i}(x;t)  \label{(3.2.16)}
\end{equation}%
We have the following immediate consequence of the methods of \cite{LvH},
Appendix A:

\begin{lemma}
\label{lem:3.2.1}In the representation defined by the state $\omega _{p,q}$,
defined by \eqref{(3.2.3)} and \eqref{(3.2.4)}, the following strong limit
exists: 
\begin{equation}
\lim_{\Lambda \nearrow \infty }\eta _{\Lambda }(\sigma ^{i};t)=M^{i}(t)%
\mathbf{1}\qquad \text{ for }i=1,2,3\   \label{(3.2.17)}
\end{equation}%
where the $M^{i}(t),i=1,2,3$, satisfy the system of nonlinear differential
equations on $\mathbb{R}^{3}$: 
\begin{equation}
\dot{M}^{1}(t)=2(b-c)M^{2}(t)M^{3}(t)  \label{(3.2.18)}
\end{equation}%
\begin{equation}
\dot{M}^{2}(t)=2(c-a)M^{1}(t)M^{3}(t)  \label{(3.2.19)}
\end{equation}%
\begin{equation}
\dot{M}^{3}(t)=2(a-b)M^{1}(t)M^{2}(t)  \label{(3.2.20)}
\end{equation}

Moreover, equation \eqref{(3.2.17)} is the precise analog of the l.h.s. of
equation \eqref{(3.2.5)} when, in $H_{\Lambda }^{\mathrm{m.f.}}$ of %
\eqref{(3.2.13)} and \eqref{(3.2.14)}, the special case $a=b=0$ is
considered.
\end{lemma}

\noindent \textit{Proof.} Concerning the mean-field model, $\omega _{p,q}$,
as defined by \eqref{(3.2.3)}, \eqref{(3.2.4)}, being a product state, i.e.,
satisfying $\omega _{p,q}(AB)=\omega _{p,q}(A)\omega _{p,q}(B)$, $\forall
A,B\in \mathcal{A}$, trivially satisfies the condition of ergodicity %
\eqref{(1.2.6)}, and therefore, by Lemma 2, p. 402 of \cite{LvH}, the strong
limits in \eqref{(3.2.17)} exist. The rest follows by the Lemma p. 435 in 
\cite{LvH}.

\hfill $\Box $

The following theorem shows that mean-field theory dramatically fails to
describe the decay in Corollary \ref{cor:3.2.1}:

\begin{theorem}
\label{th:3.2.2} Let, in \eqref{(3.2.13)}, \eqref{(3.2.14)}, 
\begin{equation}
a=b  \label{(3.2.21)}
\end{equation}%
as well as 
\begin{equation}
M^{3}(0)=0  \label{(3.2.22)}
\end{equation}%
Then, the system of nonlinear equations \eqref{(3.2.18)}, \eqref{(3.2.19)}, %
\eqref{(3.2.20)} on $\mathbb{R}^{3}$ has the unique, global solution given
by 
\begin{equation}
M^{1}(t)\equiv \pm p\qquad \text{ together with \qquad }M^{2}(t)\equiv \pm q%
\text{ \quad and \quad }M^{3}(t)\equiv 0  \label{(3.2.23)}
\end{equation}%
that is, the mean magnetization remains \textquotedblleft
frozen\textquotedblright\ at its initial value $(\pm p,\pm q,0)$. The
solution is Lyapunov stable.
\end{theorem}

\noindent \textit{Proof.} By \eqref{(3.2.21)}, \eqref{(3.2.22)} and %
\eqref{(3.2.20)}, $M^{3}(t)\equiv 0$, $\forall t\in \mathbb{R}$. Inserting
this in \eqref{(3.2.18)} and \eqref{(3.2.19)}, we obtain \eqref{(3.2.23)}.

The quantity $\Lambda(t) \equiv M^{1}(t)^{2}+M^{2}(t)^{2}+M^{3}(t)^{2}$ is a
constant of the motion (Casimir element), even without the assumption %
\eqref{(3.2.21)}, which guarantees that the system of equations %
\eqref{(3.2.18)},\eqref{(3.2.19)},\eqref{(3.2.20)} has a \emph{unique}
global solution, see \cite{LvH}, Appendix A, and references given there.

The quantity $\Lambda$ is also a Lyapunov function for the system of
nonlinear differential equations \eqref{(3.2.18)}, \eqref{(3.2.19)}, %
\eqref{(3.2.20)}, satisfying $\dot{\Lambda}(t) \equiv 0$, and therefore the
solution is Lyapunov-stable.

\hfill $\Box $

The above theorem implies, together with Lemma \ref{lem:3.2.1} and Corollary %
\ref{cor:3.2.1}, that the mean-field version of \eqref{(3.2.2)} violates
Haag's principle in the sharpest possible form, i.e., the mean magnetization
in the x-y plane remains \textquotedblleft frozen\textquotedblright\ at its
initial value, contradicting the \textquotedblleft conventional
wisdom\textquotedblright\ which views mean-field theory as the
\textquotedblleft long-range limit\textquotedblright\ of interactions of
infinite range. Thus, the phenomena of energetic and dynamic stability
violate, in general, the universality which was proved, as discussed in the
previous subsection, for the critical exponents. In particular,
nonequilibrium mean-field theory fails to describe these phenomena, by
Theorem \ref{th:3.2.2}.

\begin{remark}
\label{Remark 3.6}The rate of decay of the states $\omega _{p,q}$ also
permits to distinguish between long-range and short-range interactions: for
various short-range interactions (of exponential decrease), exact results
exist which demonstrate that the decay is polynomial (\cite{Em}, \cite{Ra}).
\end{remark}

\section{Conclusion and open problems\label{4}}

\setcounter{equation}{0} \setcounter{Thm}{0}

In \cite{Gli}, J. Glimm remarks: ``The importance of universality in
mathematics is that indicates the existence of a general theory, whose
qualitative (and quantitative) features describe a broad range of
phenomena''.

In this paper, however, we analysed some aspects of just \emph{one} universality
class, the \emph{Ising universality class} defined by the equations
\eqref{(1.2.1)}, \eqref{(1.2.2)}, and \eqref{(1.2.3)}, and most results concerned 
only the special Ising case $\delta = 0$. This class also includes Euclidean
quantum field theory \cite{GJ}.

The main objects in a future general theory might be microscopic and
macroscopic states of infinite systems (this point relates to the second
notion of phase transition discussed in Subsection \ref{1.1}), and their
mutual relationship. The \textquotedblleft discontinuities\textquotedblright
, mentioned by Glimm as the \textquotedblleft essential difference in the
analysis over finite vs. infinite dimensional spaces\textquotedblright\ (%
\cite{Gli}, p.674), which manifest themselves as singularities in the free
energy in the thermodynamic limit, might appear also, in this extended view, as
changes in basic structural properties of the states of infinite systems,
for instance in the change from pure to mixed in the theory of
irreversibility in (\cite{Wre1}, \cite{Wre2}), or in the passage from mixed
to pure states in the theory of superselection sectors originating from
interactions with the environment (\cite{NT}, \cite{NW}, see also \cite{Wre2}%
). In the present paper, the \textquotedblleft
discontinuity\textquotedblright\ also arose from taking the limit %
\eqref{(1.4.5)} in Definition \ref{Definition 1.4.1}. Indeed, starting from
a (microscopic) equilibrium (KMS) state $\omega _{T}$, the thereby defined
(macroscopic) fluctuation state $\omega _{T}^{F}$ may also exhibit basic
structural changes with regard to $\omega _{T}$.

One of the main aspects of the mutual relationship between $\omega _{T}$ and 
$\omega _{T}^{F}$ concerns the relation between the energy gap, for $T=0$,
or the rate of clustering of the ergodic components of $\omega _{T}$, for $%
0\leq T\leq T_{c}$, and certain properties of $\omega _{T}^{F}$; if there is
an energy gap at $T=0$, or the clustering is summable ($l^{1}$), at $%
0<T<T_{c}$, the fluctuations characterizing $\omega _{T}^{F}$ are normal; if
the clustering is not summable ($T=T_{c}$), $\omega _{T}^{F}$ has anomalous
fluctuations. This was shown in Theorem \ref{th:2.2.3} for $\nu =2$ and for $%
\nu =3$ under the Standard Assumption \eqref{(2.2.22)}. The above
connections refer only to the connected $2$-point Wightman functions. 

In the case of s.s.b. of a continuous symmetry (e.g., $\left\vert \delta
\right\vert =1$ in \eqref{(1.2.1)}), the absence of a gap has been proved in
the case of finite range interactions ( \cite{Wre4}, \cite{LPW}) and absence
of $l^{1}$- clustering in \cite{LPW}. Thus, Criterion A does not apply.
For long-range interactions, such as the BCS
model, s.s.b. of a continuous symmetry is compatible with an energy gap (see 
\cite{Wehrl} and references given there). 

The basic element of our framework, following previous work by one of us
(M.R.) in \cite{MR}, was to emply a smooth version of fluctuation operators,
better adapted to Fourier transform methods. Indeed, a distinction between
the critical state and the other ergodic equilibrium states is naturally
formulated for the connected two-point Wightman function, in terms of its
Fourier transform \eqref{(2.2.11b)} which is stronger than the condition of
non-$l^{1}$- clustering, but allows a considerable progress towards a proof
of the Standard Assumption (Theorem \ref{th:2.2.2}). Its analogue for the
higher ($r\geq 4$)-connected Wightman functions -- the Scaling
Assumption C (\eqref{(2.3.2)} with $\rho ^{^{\prime }}>0$ in
\eqref{(2.3.3)}) -- yields a proof of the quasi-free nature of the
fluctuation state $\omega _{T_{c}}^{F}$. Since the corresponding
equilibrium state $\omega _{T_{c}}$ is an 
interacting state, this is a third of the basic structural changes discussed
in the previous paragraph. A different structural change (see Remark \ref{Remark 2.3.2}) 
was shown to occur in the passage from finite
to infinite volume due to Proposition \ref{prop:1.5.1}, which demonstrates
that for any finite volume the fluctuation state is, for any temperature, an
interacting state. This proposition is a direct corollary of the important
work of De Coninck and Gottal \cite{dCG}, which relies on the property of
infinite divisibility of the reciprocal of the characteristic function (for
finite volume). A further consequence of their results,
Proposition~\ref{prop:1.5.2}, was used in the proof of
Theorem~\ref{th:2.3.1}.  

Several aspects of \emph{dynamics} were shown to be relevant to this general
framework. Firstly, a physically intuitive explanation of the quasi-free
property of the fluctuation states for all temperatures (given in
Remark \ref{Remark 1.4.1} was suggested by the work of Wehrl
\cite{Wehrl} on the spin 
waves in the BCS model, who showed that the fluctuation states do not
satisfy the (dynamical, equilibrium) KMS condition. Since the BCS model is a
prototype of long-range interactions, and the quasi-free property of the
fluctuation state was also proved for short-range interactions in Theorem 
\ref{th:2.3.1}, this \textquotedblleft non-equilibrium
behavior\textquotedblright\ of the fluctuations does not, apparently,
distinguish between the two main universality classes (long-range and
short-range). It is therefore natural to investigate what happens with the
other macroscopic observables -- the intensive observables -- in
non-equilibrium situations, and whether there a distinction between the two
classes may arise. We were able to investigate this question in Subsection
\ref{3.2}. The two universality classes -long-range and short-range -- are known
to exhibit different rates of decay of unstable states (see Remark
\ref{Remark 3.6}). But, in addition, a more significant aspect was
shown for the first time: in the long-range case, different rates of
fall-off of the potential yield different rates of decay, i.e.,
\textquotedblleft dynamical 
critical indices\textquotedblright\ violate universality. These results,
proved in the Dyson model in Subsection \ref{3.2} are corollaries of the
recent theorem of Albert and Kiessling \cite{AlKi} on the Cloitre function.

The Bleher-Sinai theory, via hierarchical model, is relevant to the
description of the universality classes in the Ising model. It correctly
describes the Ising model with long-range interaction and the upper critical
dimension of the Ising model \cite{ADC} and is, in fact, the only way to
understand the remarkable results on universality reviewed in
Subsection \ref{3.2}. Their nonlinear equation for the block spin
distribution does have 
a (non-unique, not normal) Gaussian solution, shown to be correct for the
Ising model with short-range interactions by Theorem \ref{th:2.3.1}. It
thereby correctly accounts for the anomalous fluctuations, the major result
of physical significance.

The following remarks concern the breakdown of universality in certain
models which are \emph{not} in the Ising universality class, but
are nevertheless very illuminating regarding the great complexity of the
universality issue.

The result of \cite{LPW} was generalized by Martin \cite{Mar} in an
optimal way to include several types of decay of the interaction
potential, in particular the case of dimension $\nu=2$, providing 
an extension of the Hohenberg-Mermin-Wagner theorem (\cite{H},
\cite{MeW}). The latter theorem precludes a first-order phase
transition (i.e., with nonzero spontaneous magnetization) in n-vector
models, with $SO(n)$ symmetry. van Enter and Shlosman \cite{vES} showed,
however, that, in a model with sufficiently nonlinear
interaction potentials which satisfies the Hohenberg-Mermin-Wagner
theorem, and therefore no spontaneous magnetization takes place, there is
a first-order transition between two \emph{non-magnetized} states.
In three or more dimensions the result is that the ordinary XY model has
a critical point separating the non-magnetized high temperature regime
from the magnetized low temperature regime: in the non-linear case this
transition is first-order.

Another non-universal example is the 3-state Potts model. in dimension two
it has an ordinary critical point. If the range is large but finite, the
transition becomes first-order. This means that the same potts permutation
symmetry is broken, both models are finite-range but the behavior is again non-universal
(\cite{GM}).

Finally, it may be therefore hoped that extensions of the
Bleher-Sinai theory to hierarchical models with continuous symmetries (see 
\cite{MCG} and references given there) may bring further insights on the
basic open problem of what should replace Criterion A in the case of models
with s.s.b. of continuous symmetries. In spite of the fact that most
magnetic materials are anisotropic, the universality class of the isotropic
model (e.g., $\delta =\pm 1$ in \eqref{(1.2.1)}) is certainly not
empty!\cite{Pel}. This is one of the major open problems, whose
satisfatory solution would be a further hint that the general theory
conjectured in this section really exists.

\appendix

\section{\textquotedblleft Ising universality class\textquotedblright }\label{A}

\setcounter{equation}{0} \setcounter{Thm}{0}

The result in this appendix provides additional support to the
expression i the title, which is
used in connection with Theorem \ref{th:1.3.1}, showing that
\emph{Ising domination} occurs in connection with the property of
existence of an energy gap. It is contained in \cite{vHBW}, but this
reference contains several (sign) misprints which may obscure
understanding, and omits certain details. We therefore include a complete proof.

Let 
\begin{equation}  \label{(A.1)}
H_{\Lambda} = -1/2 \sum_{x,y \in \Lambda} \left(J_{3}(x,y)
(\sigma_{x}^{3}\sigma_{y}^{3}-S^{2}) + H_{XY}(x,y)\right)
\end{equation}
where 
\begin{equation}  \label{(A.2)}
H_{XY}(x,y) \equiv J(x,y)
\left( \sigma_{x}^{1}\sigma_{y}^{1}+\sigma_{x}^{2}\sigma_{y}^{2} \right)
\end{equation}
and 
\begin{equation}  \label{(A.3)}
|J_{3}(x,y)| = -J_{3}(x,y) \ge |J(x,y)|
\end{equation}

Above, and in contrast with the main text, $\sigma^{i}, i=1,2,3$ will denote
the three component matrices corresponding to a general spin $S$.

The main, elementary, inequality was called in \cite{vHBW} \textquotedblleft
classical domination\textquotedblright , but we prefer to call it
\textquotedblleft Ising domination\textquotedblright , in connection with
Remark \ref{Remark 1.3.1}:

\begin{theorem}[Ising domination]
\label{thm:A.1} 
\begin{equation}
S^{2}-\sigma _{x}^{3}\sigma _{y}^{3}\geq \pm \left( \sigma _{x}^{1}\sigma
_{y}^{1}+\sigma _{x}^{2}\sigma _{y}^{2}\right)   \label{(A.4)}
\end{equation}%
i.e., the $3$-terms in \eqref{(A.1)} dominate the $XY$ terms, whether ferro-
or antiferromagnetic.
\end{theorem}

\noindent \textit{Proof.} We first take the plus sign in \eqref{(A.4)}, and
prove that 
\begin{equation}  \label{(A.5)}
S^{2} \ge \vec{\sigma}_{x} \cdot \vec{\sigma}_{y}
\end{equation}
To prove \eqref{(A.5)}, recall that $\vec{\sigma} \equiv \vec{\sigma}_{x}+%
\vec{\sigma }_{y}$ is a vector sum of two quantum spins of magnitude $S$,
and has, therefore, magnitude $L \in [0,2S]$. Equation \eqref{(A.5)} is thus
equivalent to the equation $\left|\vec{\sigma}_{x}+\vec{\sigma}_{y}\right|^{2} \le
2S(2S+1)$, but this is evident because $0 \le L \le 2S$.

In order to prove \eqref{(A.4)} with the minus sign, divide $\Lambda $ into
two sublattices, defined by the nearest neighbors of a point, and perform
the unitary transformation $\sigma _{x}^{1}\rightarrow -\sigma _{x}^{1}$ and 
$\sigma _{x}^{2}\rightarrow -\sigma _{x}^{2}$, together with $\sigma
_{x}^{3}\rightarrow \sigma _{x}^{3}$, for $y$ in one sublattice, i.e., a
rotation of $\pi $ around the $3$- axis in one sublattice. This concludes
the proof of \eqref{(A.4)}, and thereby of Theorem \ref{thm:A.1}. 

\hfill $\Box $

\begin{corollary}
\label{cor:A.2} 
Ground state and energy gap for model \eqref{(A.1)}, with general spin $S$,
long-range interactions, both signs of in the $XY$ term corresponding to
ferro or antiferromagnetic couplings, and without requiring translation
invariance. 

The tensor products of the states $|+S)_{x}$ and $|-S)_{x}$, for all $x \in
\Lambda$, are ground states of $H_{\Lambda}$, given by \eqref{(A.1)}, 
\eqref{(A.2)} and \eqref{(A.3)} with eigenvalue zero. If 
\begin{equation}  \label{(A.6)}
|J_{3}(x,y_{0})| -|J(x,y_{0})| \ge \delta >0 \text{ for all } x \in \Lambda 
\text{ and any } y_{0} \text{ fixed } \in \Lambda
\end{equation}
then, there is an \emph{energy gap} in the spectrum of the physical
Hamiltonians, i.e, any one of the (positive) Hamiltonians associated to one
of the given product states, in the thermodynamic limit.
\end{corollary}

\noindent \textit{Proof.} By \eqref{(A.1)}, \eqref{(A.2)}, \eqref{(A.3)},
and \eqref{(A.4)}, 
\begin{equation}
H_{\Lambda }\geq 1/2\sum_{x,y\in \Lambda }
\left(|J_{3}(x,y)|-|J(x,y)|\right) \left(S^{2}-\sigma _{x}^{3}\sigma
  _{y}^{3}\right)   \label{(A.7)}
\end{equation}%
The fact that the ground states are the two product states $\left\vert
+S\right) _{x}$ and $\left\vert -S\right) _{x}$, for all $x\in \Lambda $,
follows from \eqref{(A.7)}.

Consider, now, the Hamiltonian $H_{\Lambda}^{+}$, defined by imposing $|+S)_{x}$
boundary conditions at each point $x$ of the boundary of $\Lambda$.
By \eqref{(A.6)} and \eqref{(A.7)}, it follows that in the
orthogonal complement of the tensor product of the states $|+S)_{x}$, if just 
one spin is flipped (at some fixed $y_{0}$, now including the boundary)  
the following inequality holds:
\begin{equation}
H_{\Lambda }^{+}\geq \delta S>0  \label{(A.8.1)}
\end{equation}
The same result follows for the Hamiltonian $H_{\Lambda}^{-}$, defined by
imposing  $|-S_{x})$ boundary conditions:
\begin{equation}
H_{\Lambda }^{-}\geq \delta S>0  \label{(A.8.2)}
\end{equation}
Since \eqref{(A.8.1)}, \eqref{(A.8.2)} are uniform in $\Lambda $, the rest of the
assertions of Corollary \ref{cor:A.2} follows. It is to be noted that, as remarked in 
\cite{KN1}, the energy gap in the thermodynamic limit is a well-defined notion
only with respect to one of the superselection sectors, because the corresponding 
states are disjoint.

\hfill $\Box $

\begin{remark}
\label{remark:A.1}
Corollary A.2 was, for $\nu=1$, and nearest-neighbor interactions,
considerably extended by Koma, Nachtergaele, Spitzer and Starr 
(\cite{KN1}, \cite{KN2}, \cite{KNS}, \cite{NSS}) to include kink
boundary conditions and thereby proving the existence of
domains. Surprisingly, the value of the energy gap is not affected by
the existence of a domain wall, as first proved in \cite{KN1} as a  
consequence of a quantum-group $SU_{q}(2)$ - symmetry of the model.
\end{remark}

\section{Possible non Gaussian behavior at $\nu=4$}\label{B}

\setcounter{equation}{0} \setcounter{Thm}{0}

A simple general probabilistic approach has been proposed by Newman
\cite{Ne1}, of which we present a short review.

It is complementary to the present paper because it concerns the
dimension $\nu=4$ in the critical case, or $\nu=3$ in the tricritical case.

Let 
\begin{equation}
\tau _{n}^{2}\overset{\mathrm{def.}}{=}\mathrm{var}(\sigma _{\Lambda _{n}})=
\mathbb{E}\sigma _{\Lambda _{n}}^{2}-\left( \mathbb{E}\sigma _{\Lambda
_{n}}\right) ^{2}  \label{taun}
\end{equation}
be the variance of the block spin variable with respect to the equilibrium
measure \eqref{(1.1a)} with non-necessarily symmetric a priori measure $
\mu _{0}(\sigma )$ on $\mathbb{R}$ and let 
\begin{equation}
f_{n}(z)=\frac{1}{\left\vert \Lambda _{n}\right\vert }\log \mathbb{E}\exp
\left( z\sigma _{\Lambda _{n}}\right) ~.  \label{(B.2)}
\end{equation}
Tipically $\sigma _{\Lambda _{n}}=\sum_{x\in \Lambda _{n}}\sigma _{x}$
represents some extensive quantity such magnetization, energy or number of
particles contained in a finite box $\Lambda _{n}\subset \mathbb{Z}^{v}$ of
volume $\left\vert \Lambda _{n}\right\vert =n^{\nu }$ and the expectation $
\mathbb{E=E}_{\beta ,J,\Lambda _{n}}$ with respect to \eqref{(1.1a)}
omits for simplicity the dependences on its parameters and domains.
Associated with $\sigma _{\Lambda _{n}}$, $z$ is the intensive thermodynamic
variable such as external magnetic field, inverse temperature or the
negative of the chemical potential. Free or periodic conditions are imposed
in the Hamiltonian of \eqref{(1.1a)} at the boundary $\delta \Lambda
_{n}$ (with magnetic field omitted and possibly long range interactions $
J_{xy}$, $x,y\in \Lambda _{n}$). 

The following are hypotheses to be assumed accordingly:

\begin{enumerate}
\item[A.] $\exists \varepsilon >0$ so that $f_{n}(z)\longrightarrow f(z)$
exists for $z\in \left( -\varepsilon ,\varepsilon \right) $ and $f(z)$ is a
real convex function on $\left( -\varepsilon ,\varepsilon \right) $ with $
f(0)=0$;

\item[B.] $f^{\prime }(0)$ exists;

\item[C.] $\left. \mathbb{E}\sigma _{\Lambda _{n}}\right/ \left\vert \Lambda
_{n}\right\vert $ is independent of $n$;

\item[D.] $\left\vert \Lambda _{n}\right\vert \longrightarrow \infty $

\item[E.] $\sqrt{\mathrm{var}(\sigma _{\Lambda _{n}})}>0$

\item[F.] $\sqrt{\mathrm{var}(\sigma _{\Lambda _{n}})}\rightarrow \infty $

\item[H.] $f_{e}(z)\equiv f(z)+f(-z)$ satisfies 
\begin{equation*}
\sup_{z>0}\lim \sup\limits_{t\rightarrow 0}\frac{f_{e}(tz)}{z^{p}f_{e}(t)}
<\infty 
\end{equation*}
for some real number $p$;

\item[I.] $f_{n}(z)\leq f(z)$, $\forall n$ and $z\in \left( -\varepsilon
,\varepsilon \right) $.
\end{enumerate}

Weak convergence (or convergence in distribution) of a random variables
sequence $X_{n}$, $n\geq 1$, means: 
\begin{equation*}
X_{n}\overset{\mathrm{w}}{\longrightarrow }X\Longleftrightarrow \mathbb{E}
g(X_{n})\longrightarrow \mathbb{E}g(X)
\end{equation*}
for all bounded continuous functions $g$. Equivalently: $\mathbb{P}%
(X_{n}\leq \xi )\longrightarrow $ $\mathbb{P}(X\leq \xi )$, $\forall \xi \in 
\mathbb{R}$ so that $\mathbb{P}(X=\xi )=0$. Write (for $\tau _{n}>0$) 
\begin{equation}
X_{n}=\frac{\sigma _{\Lambda _{n}}-\mathbb{E}\sigma _{\Lambda _{n}}}{\tau
_{n}}\ .  \label{Xn}
\end{equation}
Since $\mathbb{E}X_{n}^{2}=1$, it follows that $\left( X_{n}\right) _{n\geq
1}$ is weakly sequentially compact hence, every subsequence of $\left(
X_{n}\right) _{n\geq 1}$ has a weakly convergent subsequence.

When B. is satisfied, $f_{\pm }(z)=f(\pm z)\mp f^{\prime }(0)z$ are
non-negative, nondecreasing function on $\left( -\varepsilon ,\varepsilon
\right) $ with $f_{\pm }(0)=0$ and the following equality $
f_{e}(z)=f_{+}(z)+f_{-}(z)$ holds.

\begin{theorem}[\textbf{main}]
\label{main}

\begin{enumerate}
\item Suppose A. B. and D. hold. Then  
\begin{equation*}
\frac{\sigma _{\Lambda _{n}}}{\left\vert \Lambda _{n}\right\vert }\overset{%
\mathrm{w}}{\longrightarrow }f^{\prime }(0)\ .
\end{equation*}

\item Suppose A., B., C., E. and I. hold. Then 
\begin{equation*}
\lim \inf_{n\rightarrow \infty }~\left\vert \Lambda _{n}\right\vert
f_{e}\left( z/\tau _{n}\right) >0\qquad \text{for any \quad }z\neq 0\ .
\end{equation*}

\item Suppose A., B., C., E., I. and 
\begin{equation}
\lim \sup_{n\rightarrow \infty }~\left\vert \Lambda _{n}\right\vert
f_{e}\left( z/\tau _{n}\right) <\infty \qquad \text{for some \quad }z\neq 0\ 
\label{lim}
\end{equation}
hold. Then 
\begin{equation*}
\lim \inf_{n\rightarrow \infty }~\left\vert \Lambda _{n}\right\vert f_{\pm
}\left( z/\tau _{n}\right) >0\qquad \text{for any \quad }z\neq 0\ .
\end{equation*}
Moreover, if $\lim \sup_{n\rightarrow \infty }$ $\tau _{n+1}/\tau
_{n}<\infty $ is assumed in addition, then 
\begin{equation*}
0<\lim \inf_{z\rightarrow 0}\frac{f_{+}(z)}{f_{-}(z)}<\lim
\sup_{z\rightarrow 0}\frac{f_{+}(z)}{f_{-}(z)}<\infty ~.
\end{equation*}

\item Suppose A., B., C., F., H. and I. hold and assume that (\ref{lim}) is
also valid. Then $p\in \left[ 1,2\right] $ and any limit $X$ of the variable 
$X_{n}$ given by (\ref{Xn}) satisfies (with $c>0$) 
\begin{equation}
\mathbb{E}X=0\qquad \text{and}\qquad \mathbb{E}X^{2}=1\ ;  \label{XX}
\end{equation}
\begin{equation}
\mathbb{E}\exp (tX)\leq \exp \left( \left\vert ct\right\vert ^{p}/p\right) \
,\qquad t\in \mathbb{R}\ ;  \label{Eexp}
\end{equation}
\begin{equation}
\mathbb{P}\left( X<-\xi \right) ,~\mathbb{P}\left( X>\xi \right) \leq \exp
\left( -\left( \xi /c\right) ^{q}/q\right) \ ,\qquad \xi \geq 0\ ,
\label{PX}
\end{equation}
where $q\in \lbrack 2,\infty )$ satisfies $1/p+1/q=1$. Thus, if $p\neq 2$, $X
$ is non-Gaussian.
\end{enumerate}
\end{theorem}

See Sec. III of \cite{Ne1} for the proof of assertions 1-4 of Theorem
\ref{main} and Sec. II of \cite{Ne1} for certain classes of models for which the
assertions 1-4 are valid.

\begin{remark}
\label{rem:B1}
Hypotheses A., B., C., F. and I., which are assumed to hold in order to
imply assertions 1-4, are in fact valid at any respectable critical point.
Hypothesis I. can actually be proven by FKG (or GKS) inequality, together
with translation invariance and reflection positivity, provided A. is
assumed valid -- the ingenious proof of Proposition 2.5 in \cite{Ne1} is
recommended to the reader.
\end{remark}
Observe that 
\begin{equation*}
\tau _{n}^{2}=\mathrm{var}(\sigma _{\Lambda _{n}})=\sum_{x,y\in \Lambda
_{n}}\left( \mathbb{E}\sigma _{x}\sigma _{y}-\mathbb{E}\sigma _{x}\mathbb{E}
\sigma _{y}\right) \ .
\end{equation*}
If 
\begin{equation}
\mathbb{E}\sigma _{x}\sigma _{y}-\mathbb{E}\sigma _{x}\mathbb{E}\sigma
_{y}=F_{n}(x-y),  \label{Fij}
\end{equation}
holds, whether by periodic conditions or FKG, then define 
\begin{equation*}
G(R)=\sum_{\left\vert x\right\vert \leq R}F_{n}(x)~.
\end{equation*}

\begin{proposition}
\label{prop:B1}Suppose (\ref{Fij}) holds with $F(x)\geq 0$ for all $x\in
\Lambda _{n}\subset \mathbb{Z}^{\nu }$ and $F(0)=0$. Then, there exist $%
c_{1},c_{2}>0$ and $K_{1},K_{2}<\infty $ such that (with $\left\vert \Lambda
_{n}\right\vert =n^{\nu }$) 
\begin{equation}
K_{1}n^{\nu }G(c_{1}n)\leq \tau _{n}^{2}\leq K_{2}n^{\nu }G(c_{2}n)
\label{KK}
\end{equation}
holds for all $n$.
\end{proposition}

The following Corollaries of assertions 1-4 of Theorem \ref{main} establish
a hyper-scaling inequality involving critical exponents and provide
sufficient conditions for the scaling limit to be non-Gaussian.

\begin{corollary}[Buckingham-Gunton inequality]
\label{cor:B.1}Let $p$ and $\eta $ be defined by 
\begin{eqnarray}
p &=&\lim \inf_{z\leftarrow 0+}\frac{\log f_{e}(z)}{\log z}  \notag \\
\frac{\nu +2-\eta }{2} &=&\lim \sup_{n\rightarrow \infty }\frac{\log \tau
_{n}}{\log n}~.  \label{indices}
\end{eqnarray}
Then $p\geq 1$ and 
\begin{equation}
2-\eta \leq \frac{2-p}{p}\nu ~.  \label{BG}
\end{equation}
\end{corollary}

\noindent \textit{Proof.} Since $f_{e}^{\prime }(0)=0$ and $\lim_{z\searrow
0}\log f_{e}(z)/\log z=zf_{e}^{\prime }(z)/f_{e}(z)=1$ by l'Hopital and
hypothesis B., there is a subsequence $z_{m}\rightarrow 0+$ satisfying $\log
f_{e}(z_{m})/\log z_{m}>1-\varepsilon $ for some $\varepsilon >0$ or,
equivalently, $f_{e}(z_{m})>z_{m}^{1-\varepsilon }$ (see the first
equation (\ref{indices})) that contradicts
$f_{e}(z_{m})/z_{m}\rightarrow 0$. So, we have $p\geq 1$. It follows
from (\ref{indices}) and assertion 2. of Theorem  
\ref{main} that
\begin{equation*}
0<\left\vert \Lambda _{n}\right\vert f_{e}\left( z/\tau _{n}\right) <n^{\nu
}\left\vert \frac{z}{\tau _{n}}\right\vert ^{p^{\prime }}<n^{\nu }\left\vert 
\frac{z}{n^{r^{\prime }}}\right\vert ^{p^{\prime }}
\end{equation*}
holds for any $p^{\prime }<p$ and $r^{\prime }<\left( \nu +2-\eta \right) /2$.
Hence 
\begin{equation}
\nu -p\frac{\nu +2-\eta }{2}\geq 0\ .  \label{nuoverp}
\end{equation}
which is equivalent to (\ref{BG}).

\hfill $\Box $

\begin{remark}
\label{rem:B.3} The critical exponent $p$ for the Ising model is $p=1+1/\delta 
$ where $\delta $ is the exponent for the magnetizations as a function of
external field at criticality: $f^{\prime }(h)\approx \left\vert
h\right\vert ^{1/\delta }$ at $\beta =\beta _{c}$, as $h\searrow 0\!$.
Equation (\ref{BG}) thus reads 
\begin{equation}
2-\eta \leq \frac{\delta -1}{\delta +1}\nu   \label{deltaeta}
\end{equation}%
The classical exponents for the Curie-Weiss model are $\delta =3$ and $\eta
=0$. Exponent $\eta =0$ holds for hierarchical models, including its $N$
vector version. Replacing the classical exponents into (\ref{deltaeta}) the
inequality turns into an equality at dimension $\nu =4$. For short-range
Ising models it is believed, and recently some of the statements have been
proved in \cite{ADC}, that a critical dimension $\nu _{c}$ ($\nu _{c}=4$)
exists such that for $\nu \geq \nu _{c}$ the scaling limit of the block spin
variable is Gaussian, with classical value exponents, and for $\nu <\nu _{c}$
the scaling limit is non-Gaussian and $\delta $ and $\eta $ may not have
their classical values. There are some examples of non-Gaussian scaling
limit for hierarchical model whose explicit construction, for instance
\cite{Fe, Zu, BAHM, MCG}, holds -- or allowed to be extended -- in
dimensions $2<\nu <4$. It would be interesting to check whether the
hyper-scaling predictions or other consequences in the following
Corollaries could be verified for such models. 
\end{remark}

\begin{corollary}
\label{cor:B.2}Suppose that $p$ and $\eta $ defined in Corollary \ref{cor:B.1}
satisfy (\ref{deltaeta}) as an equality 
\begin{equation}
2-\eta =\frac{\delta -1}{\delta +1}\nu \ ,  \label{equal}
\end{equation}%
and there are constants $\theta $ and $\psi $ such that we have 
\begin{eqnarray}
f_{e}(z) &\leq &K\left\vert z\right\vert ^{p}\left\vert \log \left\vert
z\right\vert \right\vert ^{\theta }\ ,\qquad \text{for }z\ \text{small} 
\notag \\
\tau _{n}^{2} &\geq &\varepsilon n^{\nu +2-\eta }\left\vert \log
n\right\vert ^{\psi }\ ,\qquad \text{for }n\ \text{large}  \label{lower}
\end{eqnarray}%
for some $\varepsilon $, $K\in \left( 0,\infty \right) $. Then $\psi \leq
2\theta /p$. 

Suppose that $\psi =2\theta /p$ and, in addition (with (\ref{nuoverp}) as an
equality the lower bound follows from (\ref{lower})),
\begin{equation}
\varepsilon n^{2\nu /p}\left\vert \log n\right\vert ^{2\theta /p}\leq \tau
_{n}^{2}\leq K^{\prime }n^{2\nu /p}\left\vert \log n\right\vert ^{2\theta /p}
\label{upper}
\end{equation}
for some $K^{\prime }<\infty $. Then $p=1+1/\delta \in \left[ 1,2\right] $
and any limit $X$ of $X_{n}=(\sigma _{\Lambda _{n}}-\mathbb{E}_{\Lambda
_{n}}\sigma _{\Lambda _{n}})/\tau _{n}$ in distribution satisfies (\ref{XX}),
(\ref{Eexp}) and (\ref{PX}). Hence, $X$ is non-Gaussian if $p\neq 2$. 
\end{corollary}

The next corollary is intended to deal with ordinary critical point in $\nu
=4$ dimensions and tricritical point in $\nu =6$ dimensions. It immediately
follows from Corollaries \ref{cor:B.1} and \ref{cor:B.2} that

\begin{corollary}
\label{cor:B.3} Assume that (\ref{KK}) and 
\begin{equation*}
f(h)-f^{\prime }(0)h\leq \left\{ \mfrac{K^{\prime }\left\vert h\right\vert
^{4/3}\left\vert \log \left\vert h\right\vert \right\vert ^{\theta },\qquad 
\text{for }h\ \text{small and }\nu =4}{K^{\prime }\left\vert h\right\vert
^{6/5}\left\vert \log \left\vert h\right\vert \right\vert ^{\theta ^{\prime
}},\qquad \text{for }h\ \text{small and }\nu =6} \right. \ \ 
\end{equation*}%
and 
\begin{equation*}
G(R)\geq \varepsilon R^{2}\cdot \left\{ \mfrac{\left( \log R\right) ^{\psi }\
,\qquad \text{ for large }R\text{ and }\nu =4}{\left( \log R\right) ^{\psi
^{\prime }}\ ,\qquad \text{ for large }R\text{ and }\nu =6} \right.
\end{equation*}%
hold for some $\varepsilon $, $K$, $K^{\prime }\in \left( 0,\infty \right) $
and $\theta $, $\theta ^{\prime }$, $\psi $, $\psi ^{\prime }\in \left(
-\infty ,\infty \right) $. Then, if $\psi >3\theta /2$ and $\psi ^{\prime
}>5\theta ^{\prime }/3$ any limit $X$ of $X_{n}=(\sigma _{\Lambda _{n}}-%
\mathbb{E}_{\Lambda _{n}}\sigma _{\Lambda _{n}})/\tau _{n}$ in distribution
satisfies (\ref{XX}), (\ref{Eexp}) and (\ref{PX}) for some $c>0$ with
$p=4/3$ and $q=4$ in the former and with $p=6/3$ and $q=6$ in the
latter cases.
\end{corollary}

\begin{remark}
\label{rem:B.4}
According to Lee-Yang theory as it is modernly stated in \cite{Ne4}, a
random variable $X$ is said of Newman's type $\mathcal{L}$, if  \textbf{i.}
there exist some constants $c,C>0$ such that $\left\vert \mathbb{E}\exp
(zX)\right\vert \leq C\exp \left( c\left\vert z\right\vert ^{2}\right) $
holds for all $z\in \mathbb{C}$; and \textbf{ii.} the moment generating
function $\mathbb{E}\exp (zX)$ is an even function with only pure imaginary
zeros. When $X=\sum_{i\in \Lambda _{n}}\sigma _{i}=\sigma
_{\Lambda _{n}}$, the so called Lee-Yang property, satisfied for Dyson
hierarchical Ising model \cite{HHW}, see also \cite{CE}, hierarchical $O(N)$ sigma model
\cite{Ko, Zu, Wa} and hierarchical spherical model \cite{BAHM, MCG}, in
combination with the renormalization group (RG) machinery, has allowed
to deal with both the strong and weak coupling regimes
\cite{HHW, Wa, MCG}, starting the RG dynamics from their respective
original ``a priori''  
measures, evolving towards the fixed point at the critical
point. Although the original goal of Lee-Yang program was to 
understand phase transitions directly from the distribution of
Lee-Yang zeroes, rigorous results are very few (\cite{CJN1, CJN2} and
references therein) and restrict to specific models (\cite{CJN3, MCG}
and references therein).  
\end{remark}

\begin{center}
\textbf{Acknowledgements}
\end{center}

We would like to thank Bruno Nachtergaele for correspondence on the
subject of Appendix \ref{A} and Aernout van Enter for useful remarks.

We are, above all, grateful to the referee, who advised us to eliminate
a large part of irrelevant material, and concentrate only in the Ising universality class.
This led us to a streamlined version, in which the open problems are clearly separated from
the true subject of the paper. We also thank him for several important remarks, which
contributed to render the discussion completely precise, and considerably enriched the
paper with observations, mostly regarding the concept of universality.

\end{document}